\documentclass[12pt,letterpaper]{article}
\usepackage[bf]{caption}

\usepackage{morefloats}

\usepackage[margin=1.5in, right=1.5in, top=1.5in, bottom=1.5in]{geometry}
\usepackage{layout}
\usepackage{natbib}
\bibpunct{(}{)}{;}{a}{}{,}
\usepackage{array}
\usepackage{fontenc}
\usepackage{longtable}
\usepackage{graphicx}
\usepackage{epstopdf}
\usepackage{amsmath}
\usepackage{amssymb}
\usepackage{amsthm}
\usepackage{color}
\usepackage{float}
\usepackage{epsfig}
\usepackage{epsf}
\usepackage{color}
\usepackage{rotating}
\usepackage{lscape}
\usepackage{longtable}
\usepackage{latexsym}
\usepackage[rightcaption]{sidecap}
\usepackage{changes}
\colorlet{Changes@Color}{red}
\definechangesauthor[name={mr},color=green]{*}

\usepackage{times}
\fontfamily{ptm}\selectfont

\makeatletter
\renewcommand\section{\@startsection{section}{1}{\z@}%
																		{-3.5ex \@plus -1ex \@minus -.2ex}%
																		{2.3ex \@plus.2ex}%
																		{\normalfont\large\bfseries}}
\makeatother
  \setkeys{Gin}{draft=false}

\begin{document}

\begin{center}
Evolution of Earth-like extrasolar planetary atmospheres:\\
Assessing the atmospheres and biospheres of early Earth analog planets with a
     coupled atmosphere biogeochemical model
\end{center}

\vspace*{3mm}

\begin{center}
 S. Gebauer$^{1,2}$, J. L. Grenfell$^2$, J. W. Stock$^3$, R. Lehmann$^4$, M. Godolt$^2$,\\ P. von Paris$^{5,6}$ and
          H. Rauer$^{1,2}$
\end{center}

\vspace*{3mm}

\begin{center}
(1) Zentrum f\"ur Astronomie und Astrophysik (ZAA), Technische Universit\"at Berlin (TUB), Hardenbergstr. 36, 10623 Berlin, Germany\\
\vspace*{1mm}
(2) Institut f\"ur Planetenforschung (PF), Deutsches Zentrum f\"ur Luft- und Raumfahrt (DLR), Rutherfordstr. 2, 12489 Berlin, Germany\\

(3) Instituto de Astrof$\acute{\mbox{\i}}$sica de Andaluc$\acute{\mbox{\i}}$a - CSIC, Glorieta de la Astronom$\acute{\mbox{\i}}$a s/n, 18008 Granada, Spain\\

(4) Alfred-Wegener Institut Helmholtz-Zentrum f\"ur Polar- und Meeresforschung, Telegrafenberg 43, 14473 Potsdam, Germany\\

(5) Univ. Bordeaux, LAB, UMR 5804, F-33270, Floirac, France\\

(6) CNRS, LAB, UMR 5804, F-33270, Floirac, France\\
\end{center}

\vspace*{10mm}

\begin{center}
Running title: Evolution of Earth-like atmospheres
\end{center}

\begin{center}
Corresponding author:\\ Dr. Stefanie Gebauer\\ Institut f\"ur Planetenforschung (PF)\\ Deutsches Zentrum f\"ur Luft- und Raumfahrt (DLR)\\ Rutherfordstr. 2\\ 12489 Berlin\\ Germany\\ phone: +49-(0)30-67055-454\\ e-mail: stefanie.gebauer\@ dlr.de
\end{center}

\vspace*{10mm}

\noindent "Final publication is available from Mary Ann Liebert, Inc., publishers\\ http://dx.doi.org/10.1089/ast.2015.1384"

\newpage

\begin{abstract}
\normalsize 
Understanding the evolution of Earth and potentially habitable Earth-like worlds is essential to fathom our origin in the universe. The search for Earth-like planets in the habitable zone and investigation of their atmospheres with climate and photochemical models is a central focus in exo\-planetary science.
Taking the evolution of Earth as a reference for Earth-like planets, a central scientific goal is to understand what the interactions were between atmosphere, geology and biology on Early Earth.
The Great Oxidation Event in Earth's history was certainly caused by their interplay, but the origin and controlling processes of this occurrence are not well understood, the study of which will require interdisciplinary, coupled models. In this work, we present results from our newly developed Coupled Atmosphere Biogeochemistry model in which atmospheric O$_2$ concentrations are fixed to values inferred by geological evidence. 

Applying a unique tool (Pathway Analysis Program), ours is the first quantitative analysis of catalytic cycles that governed O$_2$ in early Earth's atmosphere near the Great Oxidation Event.
Complicated oxidation pathways play a key role in destroying O$_2$, whereas in the upper atmosphere, most O$_2$ is formed abiotically via CO$_2$ photolysis.
   
The O$_2$ bistability found by \citet{goldblatt2006} is not observed in our calculations likely due to our detailed CH$_4$ oxidation scheme. We calculate increased CH$_4$ with increasing O$_2$ during the Great Oxidation Event. For a given atmospheric surface flux, different atmospheric states are possible; however, the net primary productivity of the biosphere that produces O$_2$ is unique. Mixing, CH$_4$ fluxes, ocean solubility, and mantle/crust properties strongly affect net primary productivity and surface O$_2$ fluxes. 
   
Regarding exoplanets, different "states" of O$_2$ could exist for similar biomass output. Strong geological activity could lead to false negatives for life (since our analysis suggests that reducing gases remove O$_2$ that masks its biosphere over a wide range of conditions).\\

\end{abstract}

\newpage

\section{Introduction}
In recent years more and more extrasolar terrestrial planets (mostly so-called Super-Earths with masses in the range of 1-10 Earth masses) have been detected (e.g. \citealt{leger2009,mayor2009gliese,charb2009}). Currently, we have very little knowledge about the evolutionary stages of such planets because the age of planetary systems and their host stars are not well known.
Therefore, our understanding of the evolution of Earth and its biosphere-atmosphere-surface interactions is fundamental to characterize the impact of the biosphere on such atmospheres and, hence, the resulting biosignatures\footnote{A biosignature refers to a chemical species, or set of chemical species, whose presence at suitable abundance strongly suggests a biological origin.}.  

Numerous processes - including, for example biology, radiative transfer and photochemistry, but also interactions with the surface - have a strong impact on the abundance of atmospheric molecular oxygen (O$_2$). Oxygen-bearing molecules play a
direct role in Earth's geological, geochemical, and geobiological systems. Furthermore, O$_2$ plays an indirect role in making Earth "habitable" because it leads to the formation of the biosignature molecule ozone (O$_3$), which is formed from O + O$_2$ + M $\rightarrow$ O$_3$ + M (where "M" denotes a background gas which removes excess vibrational energy) in the stratosphere. O$_3$ efficiently absorbs ultraviolet (UV) radiation which is harmful to many forms of life.  Next to other important greenhouse gases, such as water vapor (H$_2$O) and carbon dioxide (CO$_2$), O$_3$ is important for Earth's energy balance because O$_3$ (besides being radiatively active itself) and other products of O$_2$ photochemistry determine the lifetimes of atmospheric greenhouse gases, such as methane (CH$_4$), but also other trace gases such as carbon monoxide (CO) and hydrogen sulfide (H$_2$S).
Hence, the evolution of O$_2$ on Earth is one of the fundamental aspects that have determined the evolution of Earth's surface environment, climate, and biosphere.


\subsection{O$_2$ in the Modern Earth's Atmosphere}
The present-day atmosphere contains about $1.2\cdot 10^{21}$ g $=3.75\cdot 10^{19}$ mol of
O$_2$ \citep{lasaga2002}. Its abundance is regulated by the
established sources and sinks of the oxygen cycle. About 99\% of photosynthetically produced organic carbon is approximately in balance with the reverse process, namely, respiration \citep{prentice2001}, which results in an NPP of about 45 Petagrams (Pg ($10^{15}$ g)) C per year (yr) \citep{prentice2001}. A small amount of the remaining organic carbon \citep[about 0.2\%,][]{prentice2001} escapes oxidation during aerobic respiration via sedimental burial at the ocean floor, and the O$_2$ that remains is then liberated into the atmosphere (1 mol of organic carbon corresponds to 1 mol of O$_2$). This process constitutes the major source of atmospheric O$_2$. There is an additional source from the burial of other redox-sensitive chemical compounds like pyrite (FeS$_2$) and ferrous iron (FeO), whereas the burial of sulfate minerals in sediments removes a small portion of O$_2$.
Sedimental burial is estimated to account for 0.3-0.8 Pg O$_2$/yr \citep{holland2002}; this process would
require 1.5 - 4 millions of years to produce the present mass of
atmospheric O$_2$. There also exists a much smaller modern-day source
of $\sim 0.4\cdot 10^{-3}$ Pg/yr \citep{jacob1999}, which arises via H$_2$O photolysis followed by escape of atomic hydrogen from the atmosphere. Although this is only a minor
source in the modern atmosphere, it might have been much faster
for the early Earth \citep{kasting2003}. The main O$_2$
atmospheric modern-day sinks include the following: (1) weathering, estimated to remove $\sim 0.5$ Pg/yr \citep{holland2002}; (2) metamorphic reactions that occur on hot volcanic rock,
estimated to account for $\sim 0.08$ Pg/yr \citep{catling2005}
and (3) reactions with reduced
gases in the atmosphere, estimated to remove about $0.03-0.06$
Pg/yr, although this value is highly uncertain \citep{catling2005}. 


\subsection{Atmospheric O$_2$ in Earth's History}
The O$_2$ concentration has risen from its prebiotic abundance of less than $10^{-13}$ Present Atmospheric Level (PAL) \citep{kasting1981} during the Archean eon\footnote{The Archean is a geological eon starting after the Hadean eon at 4.0 Gigayears (Gyrs) ago and ending at 2.5 Gyrs ago.}, up to about 1\% PAL by about 2.3 Gyrs ago and may
have exceeded 15\% PAL by about 2.1 Gyrs \citep[e.g.,][]{knoll1995,holland2001} (see Fig. \ref{figure_sauerstoff}). Realizing the nature of the Great Oxidation Event (GOE) is central to understand the development of life on Earth and, hence, atmospheric biosignatures on Earth-like, potentially habitable extrasolar planets. Geological indices that imply an anoxic Archean atmosphere include the Mass-Independent Fractionation (MIF) of sulphur isotopes \citep [e.g.,][]{far2000,pavlov2002} 
and the existence of Banded Iron Formations (BIFs) \citep [e.g.,][]{cloud1968,holland1984,isley1999}. 
On the other hand, an oxidized Archean scenario was proposed, for example, by \citet{Ohmoto1997} based on paoleosol data. Although cyanobacteria were already present on the Earth by about 2.5 Gyrs \citep{summons1999}, for reasons not clear the recorded rise in O$_2$ took place at least 200 million years
later. The O$_2$ rise may also have changed the Earth's climate, triggering a so-called snowball Earth \citep{kopp2005} possibly due to responses in the photochemistry and/or negative impacts on methanogenic bacteria leading to a decrease in the warming due to CH$_4$ \citep[e.g.,][]{zahnle2006}. Note, however, that there are additional possible cooling mechanisms, for example, a change in volcanic gases \citep[see e.g.,][]{catling2005}. \citet{canfield2005} and \citet{holland2006} reviewed the geological evidence and current understanding of oxidative transitions in the Earth's atmosphere.
Typically, Earth system box-models (without detailed atmospheric photochemistry and climate calculations) are applied to investigate the rise in O$_2$. For example, \citet{kump2001} suggested that oxidation events in the Paleoproterozoic corresponded to a mantle overturning event, where previously oxidized mantle was brought to the surface. This would lead to less-reduced gases being emitted by volcanoes and allow atmospheric O$_2$ to rise. \citet{lasaga2002} applied a box model to investigate negative feedbacks of atmospheric O$_2$, which may account for its reasonably constant
abundance since the Phanerozoic era. For example, lowering O$_2$, according to their results, would (1) strongly lower the weathering rate and (2) increase oceanic anoxicity and, hence, increase organic carbon burial, via less oxidation of organic carbon in the ocean column. Effects (1) and (2) are both negative feedbacks that would stabilize O$_2$ in the atmosphere. \citet{holland2003}, however, questioned their model assumptions, for example, in deriving the dependence of burial
rates on oxygen abundance in seawater. \citet{catling2005} summarized the various theories proposed to account for the rise in O$_2$. Briefly, these include (1) an increase in the burial rate of organic carbon and other redox-sensitive compounds, (2) an increased supply of nutrients that stimulate photosynthesis, (3) a change in volcanic gas composition or (4) a higher hydrogen escape rate from the top of the atmosphere. \citet{claire2006} applied a zero-dimensional (0D) box-model to calculate the GOE and suggested a lowering in CH$_4$ as O$_2$ increases, whereas \citet{goldblatt2006} suggested a bistable state (oxic and non-oxic) of the atmosphere whereby O$_2$ abundances would respond very strongly to small changes in surface fluxes of O$_2$. To explain this behavior, they proposed a positive feedback
mechanism in which an initial increase in O$_2$ abundances leads to more O$_3$ and, hence, stronger UV shielding that reduces the O$_2$ photolytic sink and allows O$_2$ to build up.


\subsection{Contribution of this Work}
Until now, 0D box models were used to model Earth's atmospheric O$_2$ evolution by evaluating the important biogeochemical processes. In these studies, the atmosphere was included in a simplified way in order to investigate CH$_4$ oxidation as in the works of, for example, \citet{claire2006} and \citet{goldblatt2006}. On the contrary, there are numerical studies in which the atmospheres of early Earth and Earth-like extrasolar planets have been assessed at different time epochs by applying both one-dimensional (1D) global mean atmospheric column models as well as three-dimensional (3D) general circulation models \citep[see e.g.,][]{kienert2012,charnay2013,wolf2013,leconte2013,kunze2014,wolf2014,word2011}.

In the case of the 1D global mean atmospheric column models investigations were performed by using either climate \citep[see e.g.,][]{kasting1993,vonparis2008,goldblatt2009,kitzmann2010,word2010,kitzmann2011a,kitzmann2011b,word2013} or photochemistry \citep[see e.g.,][]{kasting1997,zahnle2006,segura2007,haqq2011} or coupled climate-photochemistry calculations \citep[see e.g.,][]{kasting1984,pavlov2001,pavlov2003,segura2003,segura2005,grenfell2007,grenfell2007cr,kaltenegger2010,rauer2011,grenfell2011,roberson2011,grenfell2012cr} thereby neglecting biogeochemical cycles. However, since biogeochemical processes strongly influence the abundance of atmospheric constituents such as O$_2$ and CO$_2$, it is important to add these processes.
Additionally, performing time-dependent integrations is a central scientific goal. However, this is currently only achievable for box-model studies \citep[e.g.,][]{claire2006,goldblatt2006}. Unfortunately, performing long integrations over millenia with state-of-the-art column models such as our scheme is not possible.

In this study, we have therefore developed a Coupled Atmosphere Biogeochemical (CAB) model with detailed photochemistry and climate calculations as well as biogeochemical surface processes that have an impact on atmospheric O$_2$. 
Therefore, the approach in our paper, which is similar to other atmospheric model studies \citep[e.g.,][]{kaltenegger2010,roberson2011}, is to "insert" Earth's evolution, for example, via O$_2$ surface concentrations from proxy data and via solar luminosity changes - as a boundary condition in the CAB model. Our main motivation thereby is to focus on understanding the associated atmospheric climatological, chemical, and biological responses.
Thus, we model the atmosphere and biosphere of early Earth analog planets consistently by including the effect of the fainter Sun, increased volcanic and metamorphic outgassing, and the impact of the biosphere on the atmosphere.
The photochemistry atmosphere module calculates atmospheric O$_2$, CO$_2$, and molecular nitrogen (N$_2$) abundances self-consistently by accounting for their in situ atmospheric sources and sinks. For the early Earth analog planetary atmosphere scenarios, we apply a snap-shot-like procedure by using the Earth system as a reference. 
We are able to reproduce the modern Earth atmospheric composition and temperature structure. We further calculate atmospheric surface fluxes required to maintain specified O$_2$, CO$_2$, and N$_2$ abundances in the
modern day atmosphere. For these surface fluxes we calculate with the biogeochemical module the corresponding NPP of the biospheres, which are needed to maintain the specified O$_2$ surface volume vmrs. 

As a case study, we investigated atmospheric chemical responses of O$_2$ in an early Earth analog atmosphere at $10^{-6}$ PAL O$_2$ by applying the Pathways Analysis Program (PAP) developed by \citet{lehmann2004}.


In the following sections, the CAB model (Section \ref{cabmodelsection}) and the PAP algorithm (Section \ref{papsection}) are decribed in detail. The CAB model is validated in Section \ref{cabvalidationsection}. Section \ref{scenariodescription} gives an overview of the simulated early-Earth analog planet scenarios whereas in Section \ref{resultssection} the results are presented and discussed. Section \ref{sensstudiessection} gives an overview of the sensitivity studies performed with the CAB model. We present our main conclusions in Section \ref{resultsection}.


\section{Coupled Atmosphere Biogeochemical (CAB) Model}\label{cabmodelsection}
\begin{figure}
\centering
\includegraphics[width=8.5cm]{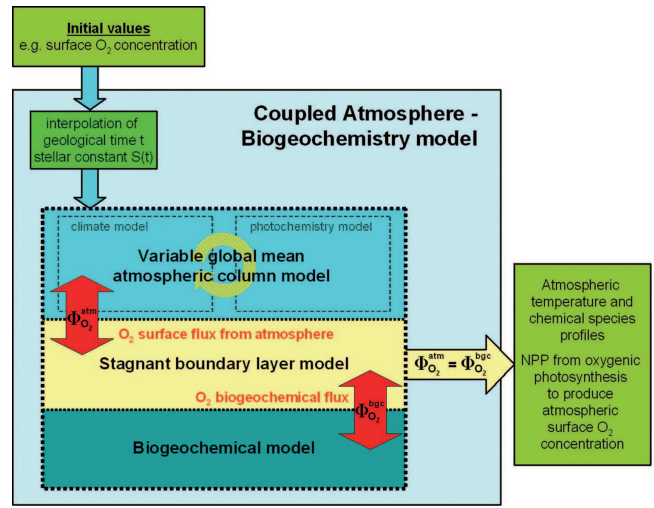} 
  \caption{Outline of the CAB model.}
  \label{figure_cabsketch}
\end{figure}

To study how the atmospheric, geological, chemical, and biotic systems interact, we have developed a Coupled Atmosphere Biogeochemical (CAB) model. It is composed of two submodules: 1. atmospheric climate and chemistry and 2. biogeochemistry. Both submodules are coupled via the stagnant boundary layer model \citep{liss1974,kharecha2005}. The overall outline of the CAB model is given in Fig. \ref{figure_cabsketch}. 
In the following sections, each submodule and their coupling will be briefly described.

\subsection{Atmosphere Module Description}\label{atmmodel}
The "variable"\footnote{Hereafter, The 1D global mean atmospheric column module is referred to as "variable" in order to distinguish the present module from previous model versions.} 1D global mean cloud-free, steady state atmospheric column module is based on the 1D model, which was described in detail by \citet{kasting1984,segura2003,grenfell2007cr,grenfell2007,rauer2011} and \citet{grenfell2011}. 

The climate module calculates temperature and water profiles from the planetary
surface up to the lower mesosphere by solving the radiative transfer
equation. Convective adjustment is applied as a common procedure (see e.g. \citet{MW1967,kasting1988venus,pavlov2000,mischna2000}) if necessary. That is, the radiative temperature gradient,$\left|\frac{dT}{dr}\right|_{rad}$, is compared to the adiabatic temperature gradient,$\left|\frac{dT}{dr}\right|_{ad}$, by calculating the convective temperature; and if the Schwarzschild criterion is fulfilled, that is, $\left|\frac{dT}{dr}\right|_{rad}>\left|\frac{dT}{dr}\right|_{ad}$, then the calculated convective temperature replaces the radiative temperature. In the troposphere a wet adiabat is calculated. The tropopause is defined as the height where both temperature profiles intersect. For the determination of the water vapor, a relative humidity distribution observed for the modern Earth from the work of \citet{MW1967} is assumed.
The radiative part of the climate
module uses a longwave radiation scheme based on the work of \citet{mlawer1997} (Rapid Radiative Transfer Model - RRTM) and a shortwave code (quadrature-$\delta$-2-stream approximation) based on the work of \citet{Toon1989}. To account for the heating and cooling effects of clouds, the model's surface albedo ($A_{\rm{surf}}$) was adjusted to 0.21 so that the converged surface temperature, $T_{\rm{surf}}$, reaches 288.15 K. 

The chemistry module features more than 200 chemical reactions and calculates concentration profiles for 54 chemical species including the hydrogen- (HO$_{\rm{x}}$), nitrogen- NO$_{\rm{x}}$, and chlorine-oxide (ClO$_{\rm{x}}$) families and sulfur-containing species, and it solves the chemical reaction network as a coupled set of continuity equations by using the backward Euler method to calculate the concentration profiles of chemical species. Below the tropopause, H$_2$O is given by the climate module, whereas above it is calculated self-consistently by the chemistry module according to its in situ atmospheric sources and sinks. The chemistry module accounts for wet deposition of chemical species as an atmospheric sink in the lower atmosphere by using solubility constants given by \citet{giorgi1985}. In the newly developed CAB model, CO$_2$ is also removed via rainout by introducing an effective Henry's law constant of $3\cdot 10^{-2}$ M atm$^{-1}$ \citep{jacob1999}. At the model boundaries, several different boundary conditions are applied. Natural biogenic and source gas emissions of CH$_4$, chloromethane (CH$_3$Cl), CO, and nitrous oxide (N$_2$O) are included at the lower boundary of the model, whereas for all other species (except O$_2$, CO$_2$, and N$_2$, see below) dry deposition is applied. At the upper boundary an effusion flux for O and CO is used.  
In the CAB model, the concentrations of O$_2$, CO$_2$, and N$_2$ are now treated as altitude dependent,that is, similar to all other chemical species and are calculated throughout the atmosphere according to their in situ chemical sources and sinks. For all scenarios, the surface volume mixing ratio (vmr) of these chemical species is held fixed, and their corresponding atmospheric surface fluxes, $\Phi_{\rm{i}}^{\rm{atm}}$, are calculated. In this work, material surface fluxes are reported as globally averaged emission rates in g/yr. After consistent calculations of gasphase chemical vertical profiles for O$_2$, CO$_2$, and N$_2$ throughout the atmosphere, these profiles are transferred to the climate part where they are interpolated to the climate grid as is already done for O$_3$, CH$_4$, H$_2$O, and N$_2$O.  
In the new variable model version, the steady-state assumption is now applied
for only two chemical species, namely atomic nitrogen (N) and excited methylene ($^1$CH$_2$), unlike previous chemistry versions of the model, which assumed
steady-state for sixteen species. These two species have extremely short lifetimes, so they require the steady-state
assumption in order to avoid numerical problems.

We have introduced further boundary conditions and input parameters that will be described in more detail in the following sections.

\subsubsection{Volcanic and Metamorphic Outgassing}\label{volcgassec}
In addition to the already existing volcanic production rates from SO$_2$ and H$_2$S in the chemistry module, we have implemented volcanic outgassing rates of the chemical species H$_2$, CH$_4$, CO, and CO$_2$. The present-day fluxes of these chemical species are summarized in Tab. \ref{tab1}.
Since the input of volcanic gases might have been higher in the past, for example, due to larger interior thermal gradients \citep{sleep1982,christensen1985} we have introduced a heat flow parameterization, $Q(t_{\rm{geo}})$, as given by \citet{claire2006} such that the production rate from volcanic outgassing equals
\begin{equation}
	P_{\rm{volc,i}}=Q(t_{\rm{geo}})\cdot P_{\rm{volc,i}}^{\rm{now}}\;\;.
\end{equation}
Here $P_{\rm{volc,i}}^{\rm{now}}$ is the present volcanic input flux for species i (given in Tab. \ref{tab1}). The heat flow $Q(t_{\rm{geo}})$ is a dimensionless analytical approximation depending on the geological time $t_{\rm{geo}}$ based on heat flow models that were derived from radioactive decay
\begin{equation}
	Q(t_{\rm{geo}}) = (4.5/(4.5 - t_{\rm{geo}}))^{\eta}
\end{equation}
where $\eta = 0.7$ according to \citet{sleep2001}. $t_{\rm{geo}}=0$ represents the present day.

The additional outgassing from metamorphic processes in the crust was introduced for H$_2$ and CH$_4$ based on the work of \citet{claire2006} and taking the crustal mineral redox buffer into account. In the photochemistry module, the metamorphic production rate is implemented for H$_2$ as
\begin{equation}\label{H2meta}
	P_{\rm{meta,H_2}}=Q(t_{\rm{geo}})\frac{\chi_{\rm{H}_2}P_{\rm{meta,C}}^{\rm{now}}}{1+\chi_{\rm{CH}_4}}
\end{equation}
and for CH$_4$ as
\begin{equation}\label{CH4meta}
	P_{\rm{meta,CH_4}}=Q(t_{\rm{geo}})\frac{\chi_{\rm{CH}_4}P_{\rm{meta,C}}^{\rm{now}}}{1+\chi_{\rm{CH}_4}}
\end{equation}
where $P_{\rm{meta,C}}^{\rm{now}}\sim 748$ Tg/yr is the present total flux of metamorphic CO$_2$ through the crust \citep{morner2002}. The parameter $\chi_{\rm{H}_2}$ is the metamorphic H$_2$/CO$_2$ speciation ratio and $\chi_{\rm{CH}_4}$ is the (CH$_4$+CO)/CO$_2$ ratio. Both values are implemented as presented in Tab. 1 in the work of \citet{claire2006} assuming equal contributions from metamorphism of graphite saturated rocks and rocks with low graphite activity. Additionally, they depend on the oxygen fugacity, $f_{\rm{O}_2}$, relative to the present quartz-fayalite-magnetite (QFM) mineral redox buffer, $\Delta f_{\rm{O}_2} = 0$, for a given crustal $p$ and $T$ conditions specified by \citet{ohmoto1977}. The oxygen fugacity is usually defined as the thermodynamic activity of oxygen in equilibrium with the rock \citep{frost1991}. Note that for simplicity, we use the QFM buffer as our standard on going back in time although the mineral assemblage might have been more reducing in the past. Nevertheless, we have also performed sensitivity runs with a more/less reducing mineral redox buffer (see Section \ref{sensstudiessection}).

 
\subsubsection{Hydrogen Escape}
In the photochemical module, we have implemented the loss of H$_2$ due to diffusion-limited escape at the upper model boundary by introducing an effusion flux for H$_2$ following \citet{walker1977} as
\begin{equation}
	\phi_{\rm{H}_2}\cong 2.5\cdot 10^{13}f_{\rm{total}}\;\;,
\end{equation}
where $f_{\rm{total}}=f_{\rm{H}_2\rm{O}}+f_{\rm{H}_2}+2f_{\rm{CH}_4}+...$ is the sum of the vmr
of all hydrogen-bearing gases above the troposphere weighted by the number of hydrogen molecules
they contain. The proportionality constant, $2.5\cdot 10^{13}$, is assumed to be insensitive to differences in temperature structure and composition between a primitive and modern atmosphere. It depends on the binary diffusion parameter between the escaping species and the background atmosphere, as well as the scale height of the background atmosphere.

\subsubsection{Young Sun Analog Star $\beta$ Com}
The stellar input spectrum for the young sun analog G0V-dwarf star $\beta$ Com \citep{ribas2005} was introduced. The spectrum was derived from observations in the UV from the International Ultraviolet Explorer (IUE) satellite archive (http://archive.stsci.edu/iue), and the synthetic NextGen model spectrum at visible and near-IR wavelengths is from the work of (Hauschildt et al., 1999), who already utilized this spectrum in their general circulation model investigating the early Earth faint young sun paradox (for further details, see e.g. Kunze et al., 2014)\nocite{kunze2014}. As for \citet{rauer2011}, the spectrum was normalized to the present solar constant of 1366 W m$^{-2}$ so that the modeled planet receives the same amount of net energy at the top-of-the-atmospere (TOA) from $\beta$ Com as the Earth from the Sun. The resulting spectrum is then scaled for the different geological times following the work of \citet{gough1981}. 


\subsection{Biogeochemical Module Description}\label{biochemdescription}
The biogeochemical module applied here is based on the approach of \citet{goldblatt2006}. Therefore, we only briefly describe the main components and processes considered. \citet{goldblatt2006} considered the atmosphere to be part of the atmosphere-ocean system and did not perform climate and photochemistry calculations. Regarding atmospheric processes, they took into account hydrogen escape and a parameterization of CH$_4$ oxidation. All atmospheric processes in their work are treated without a temperature dependance as atmospheric temperature is not calculated. Since CH$_4$ oxidation and hydrogen escape are already included in our 1D global mean atmospheric column module described in Section \ref{atmmodel} they are explicitly excluded from the biogeochemical modeling described here. We furthermore split the atmosphere-ocean system used by \citet{goldblatt2006} into the atmosphere and the ocean system. In addition to our detailed atmosphere module, described in Section \ref{atmmodel}, we focus on the marine environment in our biogeochemical module. This constitutes the strongest long-term source for atmospheric O$_2$ due to the burial of organic carbon in the ocean (continental biomass contributes only on the short-term scale and therefore, can be neglected for long-term processes). 

For the biogeochemical module, we consider two reservoirs, namely, molecular oxygen ([O$_2$]) and buried organic carbon ([C$^{\rm{org}}$]). The reservoirs are quantified in units of moles. The exchange of material fluxes between theses reservoirs are given in mol/yr and represent sources ($F_{\rm{sources}}$) or sinks ($F_{\rm{sinks}}$) for the respective reservoirs. The summation of those fluxes yields a differential equation for each reservoir according to
\begin{equation}
	\frac{d [C_i]}{dt}=F_i^{\rm{sources}}-F_i^{\rm{sinks}}
\end{equation}
where $[C_i]$ represents the considered reservoir (or chemical species). 

Several biological and non-biological processes have an effect on the O$_2$ reservoir, and these are described in the following text. The biological processes are summarized in Tab. \ref{biospheretab}. In our model, there are two types of photosynthesis - oxygenic and anoxygenic whereby organic carbon is produced without the production of oxygen.
Overall, the main context is as follows: Organic carbon produced from the different types of photosynthesis descends through the ocean column, and thereby different processes remove different fractions of the descending carbon that is finally incorporated into the ocean floor. The above is the theoretical foundation upon which the equations of \citet{goldblatt2006} are based, and this can be described in more detail as follows:
The net primary productivity (NPP) of the biosphere is described by the NPP from oxygenic photosynthesis, $N$ [mol O$_2$/yr], plus the additional input of inorganic reductants for anoxygenic photosynthesis, $r$ [mol O$_2$ equivalent/yr], which is represented by ferrous iron, Fe$^{2+}$. Therefore, the rate of total organic carbon produced by oxygenic and anoxygenic photosynthesizers in the ocean system is $(N + r)$ (see Tab. \ref{biospheretab}).
From the available organic carbon an assumed amount $\beta=0.002$ \citep{prentice2001,betts1991} is buried in sediments at the ocean floor, namely, $\beta\cdot(N + r)$ where $\beta$ is the burial efficiency. 
Heterotrophic aerobic respirers then use a fraction of organic carbon remaining, namely $\gamma=[\rm{O}_2]/(d_{\gamma}+[\rm{O}_2])$ (where $d_{\gamma}=3.7\cdot 10^{17}\mbox{ mol}$ corresponding to the inhibition of aerobic respiration at the Pasteur point, $\sim 0.01$ PAL \citep{engelhardt1974}). Afterwards, the remaining organic carbon is used by fermenters and acetotrophic methanogens that produce CH$_4$ and CO$_2$.
From the CH$_4$ produced, a certain fraction, namely $\delta=[\rm{O}_2]/(d_{\delta}+[\rm{O}_2])$, is oxidized by methanotrophs depending on the available O$_2$ concentration where $\delta_{\delta} = 2.7\cdot 10^{17}\mbox{ mol, which}$ is equivalent to $[\rm{O}_2]=2$ $\mu$M \citep{Ren1997}. M is equivalent to mol per liter. Below the critical dissolved O$_2$ concentration of 2 $\mu$M, the rate of methane oxidation is inhibited. 

\begin{figure}
\centering
\includegraphics[width=8.5cm]{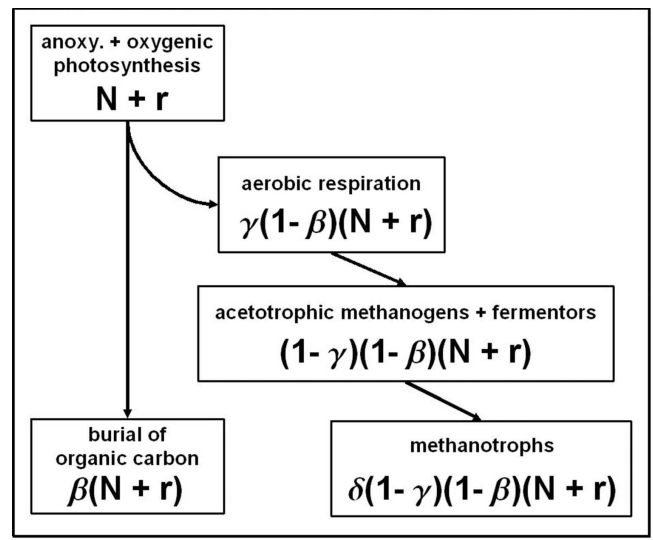} 
  \caption{Processes in the marine biosphere after \citet{goldblatt2006}. $N$ [mol O$_2$/yr] is the net primary productivity from oxygenic photosynthesis, $r$ [mol O$_2$ equivalent/yr] represents the input from anoxygenic photosynthesis. $\beta$ is the fraction of buried organic carbon. $\gamma$ resembles the fraction of aerobically respired organic carbon. $\delta$ depicts the fraction of CH$_4$ which is oxidized by methanotrophs.}
  \label{figure_bio}
\end{figure}

\begin{figure}
\centering
 \includegraphics[width=8.5cm]{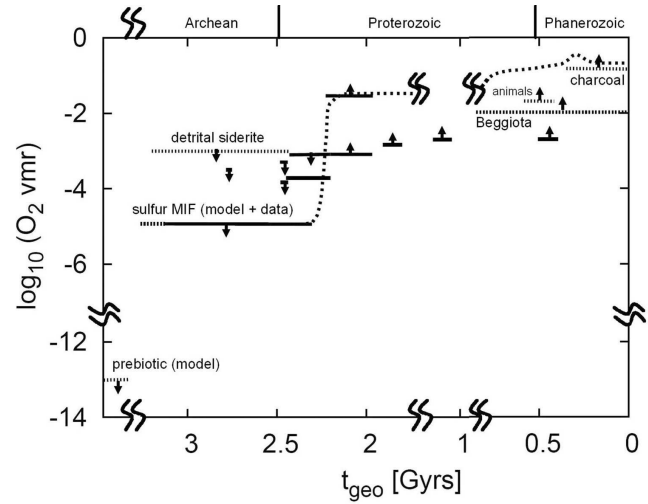} 
  \caption{A possible evolutionary path of O$_2$ in the atmosphere (thick dashed) which satisfies biogeochemical data. For further details: see text. Data are taken and modified from \citet{catling2005}.}
  \label{figure_sauerstoff}
\end{figure}

In summary, Fig. \ref{figure_bio} shows the total effect of the biosphere on the O$_2$ concentration which can be written as
\begin{equation}\label{bioO2}
	\frac{d}{dt}[\rm{O}_2]^{\rm{Bio}}=\Omega([\rm{O}_2])N - (1-\Omega([\rm{O}_2)])r + \beta(1-\Omega([\rm{O}_2]))(N+r)
\end{equation}
where $\Omega([\rm{O}_2])=(1-\gamma)(1-\delta)=(1-\frac{[\rm{O}_2]}{(d_{\gamma}+[\rm{O}_2])})(1-\frac{[\rm{O}_2]}{(d_{\delta}+[\rm{O}_2])})$ is the fraction of the organic carbon available to decomposers. For a complete derivation of this equation, we refer the reader to the work by \citet{goldblatt2006}. 

Buried organic carbon is assumed to be weathered at a rate which is determined by the rate of mechanical uplift and weathering. As in the work by \citet{goldblatt2006}, it is assumed that all organic carbon exposed by weathering is consumed by decomposers described above. The rate of weathering [C$^{\rm{org}}$] is $w$[C$^{\rm{org}}$]. 
The total effect of weathering of organic carbon for [O$_2$] is given according to \citet{goldblatt2006} by
\begin{equation}\label{o2carbweath}
	\frac{d}{dt}[\rm{O}_2]^{\rm{Cw}}=-(1-\Omega([\rm{O}_2]))w[\rm{C}^{\rm{org}}]\;\;.
\end{equation}

For the evolution of the O$_2$ reservoir in the surface ocean, Eq. \ref{bioO2} and \ref{o2carbweath} are summed up to yield
\begin{equation}\label{masterbgc}
	\frac{d}{dt}[\rm{O}_2]=\Omega([\rm{O}_2])N - (1-\Omega([\rm{O}_2]))r + (1-\Omega([\rm{O}_2]))(\beta(N+r)-w[\rm{C}^{org}])\;\;.
\end{equation}
In the same manner, we can find a differential equation for the buried organic carbon resevoir
\begin{equation}\label{masterbgcCorg}
	\frac{d}{dt}[\rm{C}^{\rm{org}}]=\beta(N+r)-w[\rm{C}^{\rm{org}}]\;\;.
\end{equation}
If we now considers steady-state conditions for the atmosphere module, then the left-hand side of Eqs. \ref{masterbgc} and \ref{masterbgcCorg} are set to zero yielding a more simplified equation, namely,
\begin{equation}\label{steadymasterbgc}
	0=\Omega([\rm{O}_2])N - (1-\Omega([\rm{O}_2]))r\;\;. 
\end{equation}
This equation is then solved for [O$_2$] using the Van Wijngaarden - Dekker - Brent root search method \citep{brent1971,press1993} (also called Brent's method) yielding the number of moles of O$_2$ depending on the input from oxygenic and anoxygenic photosynthesis $N$ and $r$ (see Section \ref{couplingsection}). This method combines linear interpolation and inverse quadratic interpolation with bisection in order to obtain the root of Eq. \ref{steadymasterbgc} efficiently. To derive the aqueous O$_2$ concentration, we calculate according to \citep{goldblatt2006}
\begin{equation}\label{aquO2}
	[\rm{O}_2]^{\rm{aq}} = \frac{H_{[\rm{O}_2]}}{\mu}[\rm{O}_2]
\end{equation}
where $\mu = 1.773\cdot 10^{20}$ mol is the number of moles in the present atmosphere \citep{goldblatt2006} and $H_{\rm{O}_2} = 0.0013$ M/atm \citep{lide1995} is the solubility constant of O$_2$ (in H$_2$O).

Note that since we set Eqs. \ref{masterbgc} and \ref{masterbgcCorg} to zero, the burial and weathering of buried organic carbon drops out of the calculations presented. To address these processes and their impact, however, the carbon cycle also has to be implemented, which will be part of future work.


\subsection{Coupling of the Atmosphere and Biogeochemical Module}\label{couplingsection}
The atmosphere and biogeochemistry modules are coupled via a so-called stagnant boundary layer model \citep{liss1974,kharecha2005}. It calculates the biogeochemical O$_2$ flux, $\Phi_{\rm{O}_2}^{\rm{bgc}}$, defined by the different processes in Section \ref{biochemdescription} from
\begin{equation}\label{fluxstag}
	\Phi_{\rm{O}_2}^{\rm{bgc}}=\rm{v}_p(\rm{O}_2)([\rm{O}_2]^{\rm{aq}}-H_{\rm{O}_2} p_{\rm{O}_2})C
\end{equation}
where $\rm{v}_p(\rm{O}_2)$ is the piston velocity of O$_2$, $p_{\rm{O}_2}$ is the atmospheric partial pressure of O$_2$, and $C=6.02\cdot10^{20}$ molecules cm$^{-3}$/mol l$^{-1}$ is a units conversion factor.
The piston velocity is a measure of the diffusitivity of a certain chemical species through a stagnant layer. For a chemical species i it is defined as $\rm{v}_p(i)=K_{\rm{diff}}(i)/z_{\rm{film}}$ with the thermal diffusivity, $K_{\rm{diff}}(i)$ and the thickness of the stagnant layer, $z_{\rm{film}}$ (here $\approx 40$ $\mu$m as in the work of \citet{kharecha2005}).
For O$_2$, $\rm{v}_p(\rm{O}_2)$ is derived according to \citet{wilke1955} resulting in a value of $\rm{v}_p(\rm{O}_2)\approx 5.7\cdot 10^{-3}$ cm s$^{-1}$ at 25 $^{\circ}$C. 

A solution of the whole CAB model is found if 
\begin{equation}\label{2fluxes}
	\Phi_{\rm{O}_2}^{\rm{atm}}-\Phi_{\rm{O}_2}^{\rm{bgc}}=0
\end{equation}
where $\Phi_{\rm{O}_2}^{\rm{atm}}$ is calculated by the atmosphere module for a given surface vmr. Eq. \ref{2fluxes} is also solved by applying the Van Wijngaarden - Dekker - Brent root search method. In this work, $r$ is held constant at the modern-day input of reductants into the ocean, and $N$ is varied stepwise according to the applied root search method between $N_{low}=1$ mol O$_2$/yr and $N_{up}=1\cdot 10^{25}$ mol O$_2$/yr in order to determine the input from oxygenic photosynthesis $N$ into the atmosphere that is required to maintain the prescribed atmospheric conditions.

\section{Pathways Analysis Program}\label{papsection}

The Pathway Analysis Program (PAP) was developed by \citet{lehmann2004} to automatically identify chemical pathways in arbitrary chemical reaction networks and to quantify their efficiencies by assigning rates. PAP was applied by, for example, \citet{grenfell2006,verronen2011,stock2012a,stock2012b,grenfell2013} and \citet{verronen2013}. The algorithm yields a list of all dominant pathways that produce, destroy, or recycle a chemical species of interest.
PAP requires as input a complete list of chemical species and their concentrations, as well as their concentration changes caused by chemical reactions during a specified time interval. Furthermore, a complete list of reactions and their corresponding rates is required. Starting with individual reactions as initial pathways, longer pathways are formed step-by-step. For this, shorter pathways that have already been found are connected at so-called "branching point" species, whereby each pathway that forms a branching point species is connected with each pathway that destroys it. Branching point species are chosen based on increasing lifetime with respect to the pathways constructed so far. In this work, all chemical species with a chemical lifetime shorter than that of O$_2$ are treated as branching point species.
Since, in general, the chemical lifetime of a chemical species varies with altitude, the choice of branching point species adapts to the local chemical and physical conditions. 

For the analysis of large and complex reaction networks, chemical pathways with a rate below a user-defined threshold rate $f_{\rm{min}}$ are deleted to avoid long computational time. In the present study, $f_{\rm{min}}=10^{-13}$ parts per billion by volume per second (ppbv/s)) was chosen, which is sufficient for finding all dominant pathways producing or consuming atmospheric O$_2$. A PAP analysis was performed for each of the 64 vertical atmospheric column module chemistry layers. The resulting production and destruction rates of O$_2$ from each individual pathway are integrated over the atmospheric module vertical grid and are expressed as a percentage of the total column-integrated production and destruction rate from pathways found by PAP.

\section{CAB Model Validation}\label{cabvalidationsection}
\subsection{Atmosphere Module}
The CAB model was applied to modern Earth conditions to compare the modified atmosphere module presented in this work to the previous atmosphere model version discussed by\citet{rauer2011} (denoted as RG2011), which was validated against modern Earth. Results suggest that over the calculated pressure (altitude) range a maximum relative change of the temperature profile against RG2011 between -0.5\% to 0.1\% is observed. The surface temperatures deviate by 0.02 K. Hence, an overall good agreement of the temperature profiles between both models was obtained. 

Tab. \ref{table:2} shows the maximum relative change in the atmosphere of relevant chemical species vmr profiles against RG2011 over altitude for O$_3$, H$_2$O, CH$_4$, N$_2$O, CH$_3$Cl, CO$_2$. For O$_2$ and N$_2$ only small deviations are obtained. 
Due to the introduced hydrogen escape at the upper boundary, a maximum relative change of -90\% for H$_2$ is found (not shown) at the TOA, whereas below 40 km the change is negligible. Quite strong changes occur for sulfur-bearing species, for example, H$_2$S (-64\%), SO$_2$ (-73\%), SO (-72\%) (not shown), because updated volcanic SO$_2$ and H$_2$S emissions were implemented, and furthermore, new volcanic emissions for H$_2$, CO, CH$_4$, and CO$_2$ were introduced into the CAB model (see Tab. \ref{tab1}). Despite these changes, the biosignature species, O$_3$ and N$_2$O, and biosignature related compounds, CH$_4$ and H$_2$O, compare very well with the previous global atmospheric column model used by RG2011, which is shown in Tab. \ref{table:3}.

\subsection{Biogeochemical Module}
For modern Earth conditions and the present input of inorganic reductants (Fe$^{2+}$) into the ocean of $r=r_0=3\cdot 10^{11}$ mol Fe/yr \citep{holland2006}, which is equal to $r_0=7.5\cdot 10^{10}$ mol O$_2$ equivalents/yr, the CAB model calculates a global net primary productivity from oxygenic photosynthesis of $N=1.05$ Petamoles (Pmol ($10^{15}$)) O$_2$/yr, which is of the same magnitude as the present observed value of $N_0\approx 3.75$ Pmol O$_2$/yr \citep{prentice2001}. The range of observed oceanic NPP varies from 2.067 - 4.167 Pmol/yr \citep{woodwell1978,antoine1996,longhurst1995}, hence the lower value matches our model value within a factor of 2.
The burial rate of organic carbon, $\beta(N+r)$ (see Fig. \ref{figure_bio}), calculated by the CAB model is 2.1 Tmol/yr. This value is in the range of the observed value of $10\pm3.3$ Tmol/yr \citep{holland1978}. Global mean burial rates, are not well known, being based on measurements at individual sites. It is particularly challenging to estimate associated uncertainties,for example, the proxy data is severely limited in time and space and, hence, difficult to estimate in a global model.


\section{Application to Early Earth Analog Atmospheres}\label{scenariodescription}
The CAB model is applied to early Earth analog atmosphere scenarios before, during, and after the GOE at about 2.3 billion years ago. For all runs belonging to the control scenario, we assume:
\begin{itemize}
	\item solar input spectrum \citep{Gueymard2004},
	\item surface pressure $p_{\rm{surf}}=1$ bar \citep [as suggested by, e.g.,][]{Som2012},
	\item gravity acceleration $g=981$ cm s$^{-2}$,
	\item biogenic surface fluxes of CH$_4$ (474 Tg/yr), CO (1796 Tg/yr), N$_2$O (13.5 Tg/yr), CH$_3$Cl (3.4 Tg/yr) (these values are set to modern Earth conditions due to missing data), H$_2$ deposition velocity ($v_{\rm{dep}} = 2.6\cdot 10^{-3}$ cm s$^{-1}$), and deposition velocities of other chemical compounds as in RG2011,
	\item crustal mineral redox buffer is set to QFM ($\Delta f_{\rm{O}_2}=0$),
	\item surface vmr of CO$_2$ of 355 ppm \citep [e.g.,][suggested only a moderate CO$_2$ increase of up to 3 PAL CO$_2$]{rosing2010}.
	\end{itemize}

We use the evolutionary path of O$_2$ by \citet{catling2005} as an input parameter for the O$_2$ surface vmr (see Tab. \ref{runtable}), which is shown in Fig. \ref{figure_sauerstoff} as a thick dashed line. It is constrained from upper and lower biogeochemical limits \citep [for further details see the work of][]{catling2005}. Dotted horizontal lines with upper bounds (downward arrows) and lower bounds (upward arrows) show biogeochemical constraints on the O$_2$ partial pressure ($p_{\rm{O}_2}$). The existence of paleosols \citep{Rye1998} are indicated by unlabeled solid horizontal lines. O$_2$ concentrations in the prebiotic atmosphere at 4.4 Gyrs are based on numerical calculations by \citet{kasting1993early}. From the observations of MIFs in pre-2.4 Gyrs sulfur isotopes, photochemical model results of \citet{pavlov2002} constrain the O$_2$ vmrs before 2.3 Gyrs. A lower limit for $p_{\rm{O}_2}$ is suggested based on the O$_2$ requirements of: (1) Beggiatoa\footnote{Beggiatoa is a colorless, filamentous proteobacteria which oxidizes H$_2$S.} emerging after 0.8 Gyrs \citep{Canfield1996}; (2) animals appearing after 0.59 Gyrs \citep{Runnegar1991}; and (3) charcoal production occuring after 0.35 Gyrs before present \citep{Chaloner1989}. The high O$_2$ concentration around 300 Ma is based on calculations by \citet{Berner2000}.
We then interpolate the corresponding geological time $t_{\rm{geo}}$ given in Gyrs for a given O$_2$ concentration and the stellar constant $S(t_{\rm{geo}})$ is then derived from the work of \citet{gough1981} according to $t_{\rm{geo}}$.

The following boundary conditions are additionally varied according to the geological time $t_{\rm{geo}}$:
\begin{itemize}
	\item volcanic emissions of H$_2$, H$_2$S, SO$_2$, CO, CH$_4$, and CO$_2$ and metamorphic emissions of H$_2$ and CH$_4$ (following Section \ref{volcgassec}).
\end{itemize}


\section{Results}\label{resultssection}
\subsection{Atmosphere Modeling}
\subsubsection{Climate Responses}

\begin{figure}
\centering
\includegraphics[width=8.5cm]{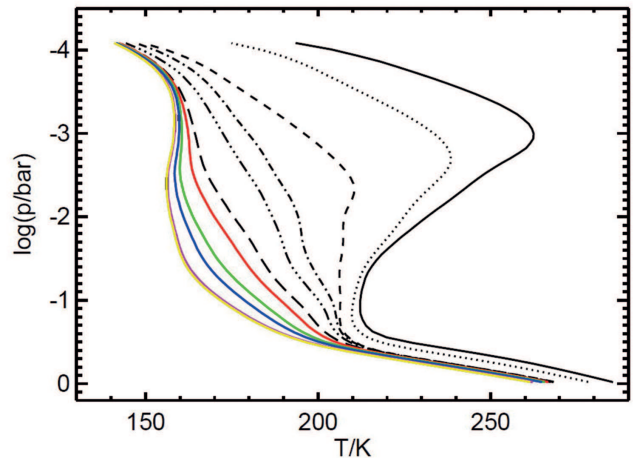} 
  \caption{Selected p-T profiles of early Earth analog atmosphere control scenario runs calculated by the CAB model. Notation: solid black: 1 PAL O$_2$ (modern Earth), dotted: $2.5\cdot 10^{-1}$ PAL ($t_{\rm{geo}}=0.82$ Gyrs), short dashed: $10^{-1}$ PAL ($t_{\rm{geo}}=2.18$ Gyrs), dash-dot: $2.5\cdot 10^{-2}$ PAL ($t_{\rm{geo}}=2.22$ Gyrs), dash-dot-dot-dot: $10^{-2}$ PAL ($t_{\rm{geo}}=2.22$ Gyrs), long dashed: $2.5\cdot 10^{-3}$ PAL ($t_{\rm{geo}}=2.23$ Gyrs), red: $10^{-3}$ PAL ($t_{\rm{geo}}=2.24$ Gyrs), green: $2.5\cdot 10^{-4}$ PAL ($t_{\rm{geo}}=2.25$ Gyrs), blue: $10^{-4}$ PAL ($t_{\rm{geo}}=2.28$ Gyrs), magenta: $10^{-5}$ PAL ($t_{\rm{geo}}=2.59$ Gyrs), yellow: $10^{-6}$ PAL ($t_{\rm{geo}}=2.69$ Gyrs).}
  \label{figure_pTvarO2}
\end{figure}

Fig. \ref{figure_pTvarO2} shows the temperature profiles of a selected number of calculated early Earth analog atmosphere runs in comparison to the modern Earth profile calculated by the CAB model.
For the modern Earth p-T profile ($t_{\rm{geo}}=0$ Gyrs, solid black) the temperature decreases with decreasing pressure from $T_{\rm{surf}}=288$ K at the surface to about $T\approx212$ K at the tropopause at $p\approx0.1$ bar. The distinct temperature maximum at the stratopause is caused by radiative heating due to the absorption of UV radiation by O$_3$. At the stratopause with a pressure of $p\approx10^{-3}$ bar a maximum temperature of $T\approx262$ K is reached. 

\begin{figure}
\centering
\includegraphics[width=8.5cm]{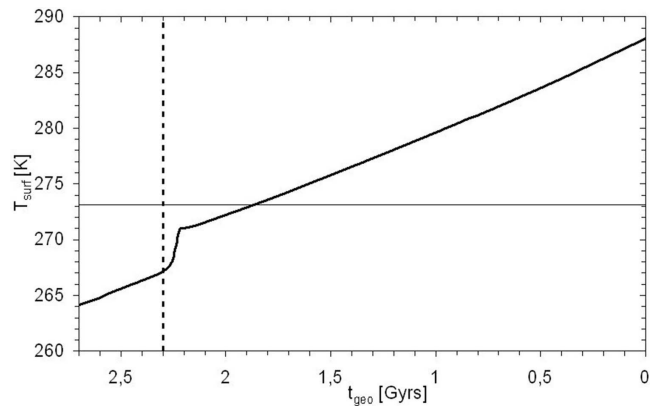} 
  \caption{Surface temperature for calculated Earth-in-time atmosphere runs. Additionally indicated as a horizontal line is the freezing temperature of H$_2$O of $T=273.15$ K. The beginning of the GOE at $t_{\rm{geo}}=2.3$ Gyrs is indicated as a vertical dashed line.}
  \label{figure_Tdecrease}
\end{figure}

Going forwards in geological time $t_{\rm{geo}}$, the O$_2$ content, thereby also O$_3$, and the solar radiation input at the TOA increase. For the $10^{-6}$ PAL O$_2$ atmosphere run at $t_{\rm{geo}}=2.689$ Gyrs ago, the solar flux incident at the TOA was decreased to about 81\% of the modern Earth value ("faint young Sun"). Furthermore in the low O$_2$ atmosphere the incoming radiation is less absorbed by molecules such as O$_3$ (total column: 15.8\% of downwards shortwave radiation is absorbed compared to 25.5\% for modern Earth) due to smaller atmospheric chemical abundances of O$_3$ (see Fig. \ref{figure_O3varO2}) and other greenhouse gases. A diminishing of the temperature inversion is observed for the low O$_2$ atmosphere runs as a result of the low O$_3$ concentrations (see Fig. \ref{figure_O3varO2}). Altogether, this leads to an increase in surface temperature (see Fig. \ref{figure_Tdecrease}) by 24 K (2.689 Gyrs ago at $10^{-6}$ PAL O$_2$: $T_{\rm{surf}}=264$ K, today at 1 PAL O$_2$: $T_{\rm{surf}}=288$ K). Note that in the case of the low O$_2$ atmosphere runs before 1.85 Gyrs ago, global surface temperatures are below 0$^{\circ}$C implying no habitable surface conditions. Results in Fig. \ref{figure_pTvarO2} are only applicable for modern Earth CO$_2$ concentrations since we have kept CO$_2$ at modern Earth concentrations in order to analyze the atmospheric responses on reducing surface O$_2$ concentrations firstly. However, see also sensitivity run G in Section \ref{sensstudiessection} with 5 PAL CO$_2$.
Further note, however, that it was recently shown by 3D general circulation models used to investigate the early Earth, for example, in the works of \citet{wolf2013,kunze2014}, that, even with global mean surface temperatures below the freezing point of water, areas with liquid surface ocean water could still exist, which implies habitable conditions.

Since the thermal radiation scheme RRTM is only valid for atmospheres that are quite close to modern Earth conditions \citep [see e.g.,][]{segura2003,vonparis2008,rauer2011}, we have compared the thermal fluxes of RRTM with thermal fluxes computed by the
line-by-line radiative transfer code SQuIRRL \citep [Schwarz\-schild Qua\-dra\-ture Infra\-Red Radiation Line-by-line,][]{schreier2001,schreier2003}.
For all runs considered the maximum relative change of net TOA fluxes calculated by RRTM vs. SQuIRRL amounted to a maximum value of -1.4 \%.


\subsubsection{Photochemistry Responses}\label{photochem}

\paragraph{\textbf{UV environment}}
To a large extent, atmospheric chemistry is driven by the incoming TOA UV fluxes. Too understand the absorbing nature of the atmosphere, UVA (315-400 nm), UVB (280-315 nm) and UVC (176-280 nm) radiation was calculated at the surface for reduced O$_2$ concentration and incoming TOA solar flux. UVA and UVB wavelength ranges are taken from the International Standard (ISO 21348) 2007. Note that in the photochemistry module the term "surface" denotes the lowermost layer at a height of about 500 m for modern Earth. In the case of the 10$^{-6}$ PAL O$_2$ atmosphere, the lowermost layer sinks to a height of $z_1=375$ m.
UVA radiation at the surface stays more or less constant for atmospheres before 2.18 Gyrs. It increases from about 75 W m$^{-2}$ at $t_{\rm{geo}}=2.18$ Gyrs ago (0.1 PAL O$_2$) to about 90 W m$^{-2}$ for modern Earth.
For the considered runs, a steady decrease from more than 12 W m$^{-2}$ to 2.3 W m$^{-2}$ is observed for surface UVB radiation before the GOE and after the GOE, respectively. The modern Earth control run exhibits a surface UVB radiation of 2.3 W m$^{-2}$, which is broadly similar to the observed value for cloud-free conditions of 1.4 W m$^{-2}$ \citep{wang2000}.
Going forward in geological time, UVC radiation decreases from 2.8 W m$^{-2}$ before 2.7 Gyrs to virtually zero $t_{\rm{geo}}=2.25$ Gyrs ago. 
This behavior is due to increasing O$_2$ and, hence, O$_3$ and H$_2$O concentrations that result in more absorption of photons in the atmosphere and more shielding of the surface from UV radiation. The impact on UVC is directly correlated with the increase in O$_3$, whereas UVB is affected by O$_3$ and H$_2$O together.

\begin{figure}
\centering
\includegraphics[width=8.5cm]{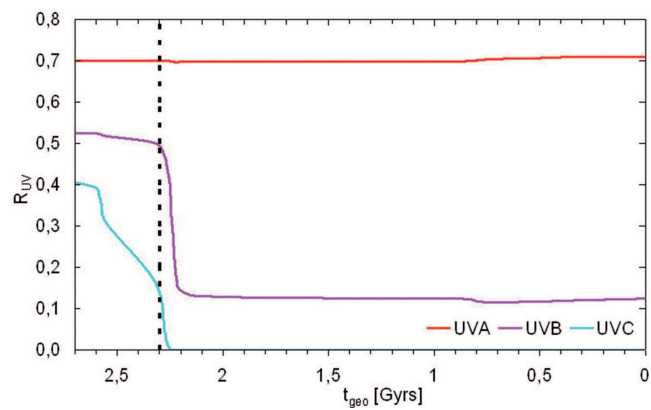} 
  \caption{Ratio (surface/TOA) radiation fluxes, $R_{\rm{UV}}$, for UVA (red), UVB (magenta) and UVC (cyan) as a function of geological time $t_{\rm{geo}}$. The beginning of the GOE at $t_{\rm{geo}}=2.3$ Gyrs is indicated as a vertical dashed line.}
  \label{figure_ruv}
  \end{figure}

\begin{figure}
\centering
\includegraphics[width=8.5cm]{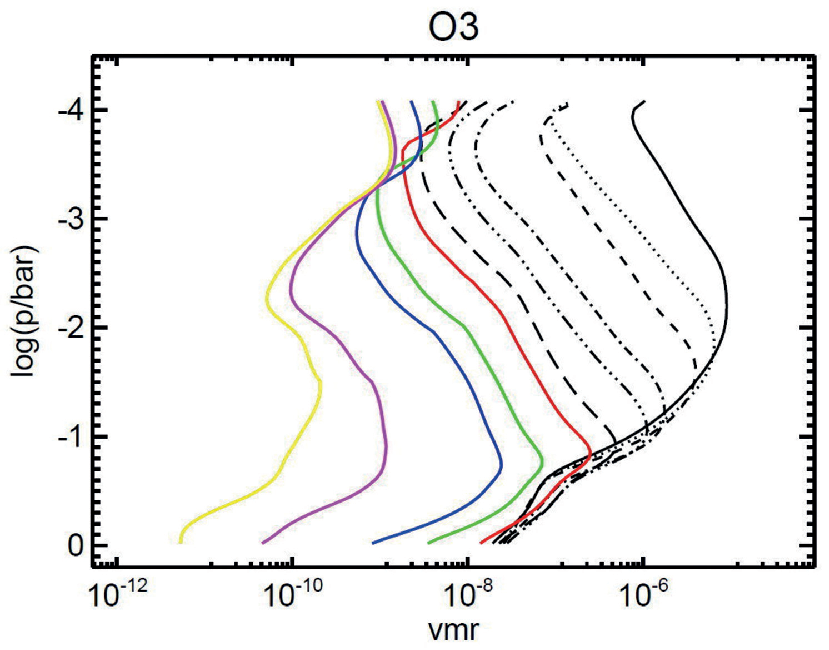} 
  \caption{O$_3$ profiles of Earth-in-time atmosphere runs calculated by the CAB model. Notation as in Fig. \ref{figure_pTvarO2}.}
  \label{figure_O3varO2}
\end{figure}

Fig. \ref{figure_ruv} shows the ratio surface/TOA radiation, $R_{\rm{UV}}$, as a measure of the radiation shielding efficiency of an atmosphere for UVA, UVB, and UVC radiation on going forward in geological time.
For high $R_{\rm{UV}}$ values, the radiation passes efficiently through the atmosphere, whereas for $R_{\rm{UV}} = 0$ it is totally blocked. 
UVA radiation passes efficiently through the atmosphere for all calculated runs. 
With the beginning of the GOE, UVB radiation is blocked rather strongly by the atmosphere, whereas
UVC radiation is already starting to be blocked 2.6 Ga ago. After the GOE the atmosphere is totally opaque for UVC radiation.
Possible negative impacts on organisms in the marine environment before the GOE might have been reduced due to the Urey effect \citep{urey1959} where a small quantity of O$_2$ produced during photosynthesis of H$_2$O absorbs a considerable amount of UV radiation. For modern Earth, UVB radiation is almost totally blocked by the atmosphere.

In the following photochemical analysis, we only focus on the impact on important atmospheric constituents such as biosignatures (O$_3$, N$_2$O, O$_2$) and related bioindicators (CH$_4$, H$_2$O) as well as CO$_2$. "Bioindicators" are indicative of biological processes but can also be produced abiotically.
\paragraph{\textbf{Ozone - O$_3$}}
O$_3$ is suggested to be a biosignature because it is formed in the stratosphere mainly from molecular oxygen\footnote{On modern Earth O$_2$ is almost exclusively produced from oxygenic photosynthesis and consumed by respiration on the short-term scale. On the long-term scale O$_2$ accumulates in the atmosphere via the burial of photosynthetically produced organic carbon and other redox-sensitive compounds.} via the Chapman mechanism \citep{chapman1930}, which is initiated by the photolysis of mostly biogenic O$_2$. In the troposphere O$_3$ is produced via the smog mechanism \citep{haagen1952}, which requires O$_2$, volatile organic compounds, nitrogen oxides, and UV radiation. The destruction of O$_3$ in the stratosphere proceeds via catalytic cycles that involve, for example, HO$_{\rm{x}}$, NO$_{\rm{x}}$, or ClO$_{\rm{x}}$ (e.g., \citealt{bates1950,crutzen1970}), which are stored in reservoir species and can be activated by changes in, for example, temperature or UV radiation. In the troposphere, O$_3$ is removed via wet/ dry deposition, photolysis, or reactions with NO$_{\rm{x}}$, HO$_{\rm{x}}$ and unsaturated hydrocarbons.

Fig. \ref{figure_O3varO2} shows the O$_3$ vmr profiles for reduced O$_2$ content of the atmosphere. Modern Earth's O$_3$ layer (solid black) exhibits a distinct maximum at about 0.01 bar due to the mechanisms decribed above. Fig. \ref{figure_O3varO2} suggests that, for reduced O$_2$ concentrations, the O$_3$ layer moves to higher pressures (lower altitudes), for example, for 10$^{-5}$ PAL to 0.2 bar, because in consequence of a lower O$_2$ concentration, less O$_3$ is produced at the same altitude. This results in increased UV radiation in atmospheric layers directly below, which stimulates the Chapman mechanism and produces more O$_3$, and hence the peak of the O$_3$ layer moves to lower altitudes.
However, if O$_2$ concentrations fall below those of CO$_2$ ($=3.6\cdot 10^{-4}$, left of the green line), results are consistent with CO$_2$ photolysis now being the major source for atomic oxygen (O) and, hence, O$_3$. The O$_3$ peak moves upwards where CO$_2$ is photolyzed more easily. Additionally, in the upper atmosphere a second O$_3$ maximum becomes visible, which is related to an increase in O$_2$ vmr toward lower pressures (see Fig. \ref{O2varO2fig}). Note that for modern Earth this secondary O$_3$ maximum lies in the vicinity of the mesopause \citep [see e.g.,][]{evans1968,hays1973,smith2005} and arises due to the interplay between active-hydrogen and active-oxygen chemistry, relatively low temperatures, and a local maximum of O \citep{allen1984,smith2005}.

\begin{figure}  
 \centering
\includegraphics[width=8.5cm]{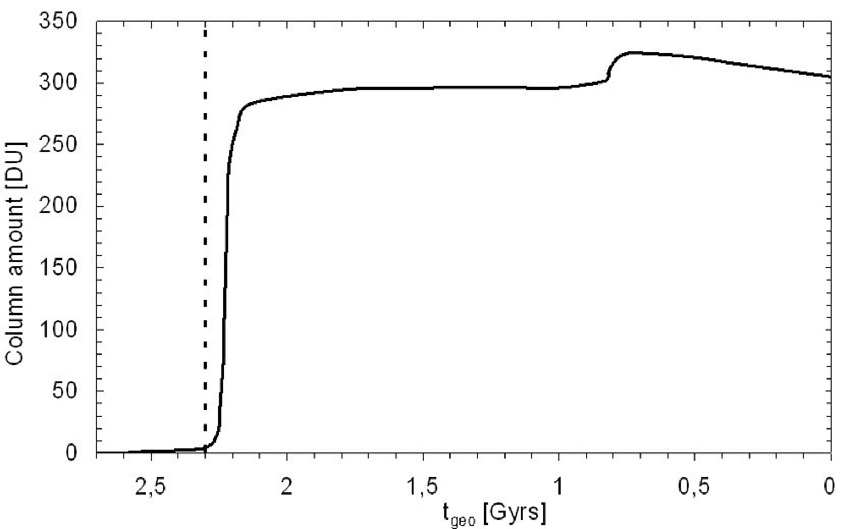} 
  \caption{O$_3$ column [DU] amount as a function of geological time $t_{\rm{geo}}$ for considered Earth-in-time atmosphere runs. The beginning of the GOE at $t_{\rm{geo}}=2.3$ Gyrs is indicated as a vertical dashed line.}
  \label{figure_O3column}  
\end{figure}

\begin{figure}
\centering
\includegraphics[width=8.5cm]{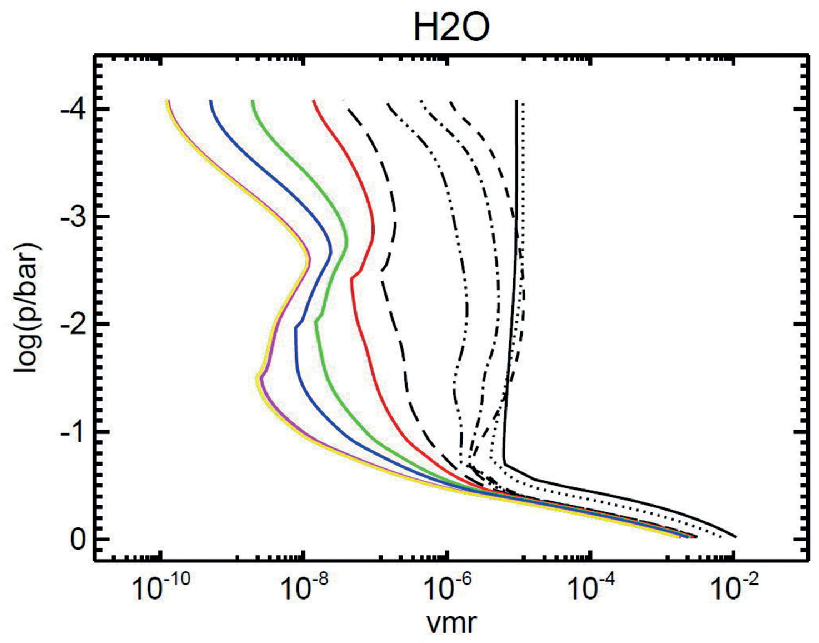} 
  \caption{H$_2$O profiles of Earth-in-time atmosphere runs calculated by the CAB model. Notation as in Fig. \ref{figure_pTvarO2}.}
  \label{H2OvarO2fig}
\end{figure}

The change in overall O$_3$ column amount on going forwards in geological time is shown in Fig. \ref{figure_O3column}. It increases from zero before the GOE to 296 DU after the GOE. At about 0.8 Gyrs ago, a Second Oxidation Event (SOE) occured resulting in an increase of the O$_3$ column amount to 325 DU. Afterward, it decreases to the modern Earth value of 305 DU. To illustrate this decrease in O$_3$, we have included Fig. \ref{H2OvarO2fig}, in which it can be seen that the troposphere becomes damper (going forward in time) due to evaporation because of increasing surface temperatures. However, in the stratosphere this is not always the case (compare the solid black line with the short-dashed and dotted line). A dryer stratosphere leads to less HO$_{\rm{x}}$ and, hence, enhanced NO$_{\rm{x}}$ resulting in an increase in catalytic O$_3$ loss. The dryer stratosphere is a result of a decrease in CH$_4$ (see Fig. \ref{CH4varO2fig}) that is driven by OH, which has a complex photochemistry.
\citet{berkner1965} estimated that effective UV shielding would be provided by a column amount of 200 DU, which is reached in this work at $t_{\rm{geo}}=2.22$ Gyrs ago (0.025 PAL O$_2$), whereas \citet{ratner1972} adopted a more stringent lower limit on the O$_3$ column amount of 259 DU (reached in this work at $t_{\rm{geo}}=2.19$ Gyrs ago (0.1 PAL O$_2$)). 


\paragraph{\textbf{Water - H$_2$O}}
In the case of modern Earth, the water vapor vmr in the troposphere is chemically inert and subject to the hydrological cycle. Above the tropopause, however, the H$_2$O vapor concentration is determined by its chemical sources (here, CH$_4$ oxidation) and sinks as well as transport from the troposphere and approaches an isoprofile for the modern Earth.

Fig. \ref{H2OvarO2fig} shows the H$_2$O vapor vmr profiles on decreasing ground level O$_2$ concentrations on going back in time. The solar flux incident at the TOA decreases, which results in a decrease in surface temperature and, hence, condensation of H$_2$O in the troposphere. The atmosphere becomes dryer due to decreasing surface temperatures. The stratospheric H$_2$O profile deviates from the isoprofile (found for modern Earth), and a distinct maximum develops in the early Earth analog runs. 
In this region, H$_2$O is primarily produced from O$_2$ (which increases in this region - see later) via the net reaction CH$_4$ + O$_2$ $\rightarrow$ H$_2$O + H$_2$ + CO which is initiated by the photolysis of CH$_4$.
Furthermore, H$_2$O is predominantly destroyed via photolysis in the low O$_2$ atmosphere due to enhanced UV radiation resulting in the formation of H$_2$ and CO$_2$. 

\begin{figure}
\centering
\includegraphics[width=8.5cm]{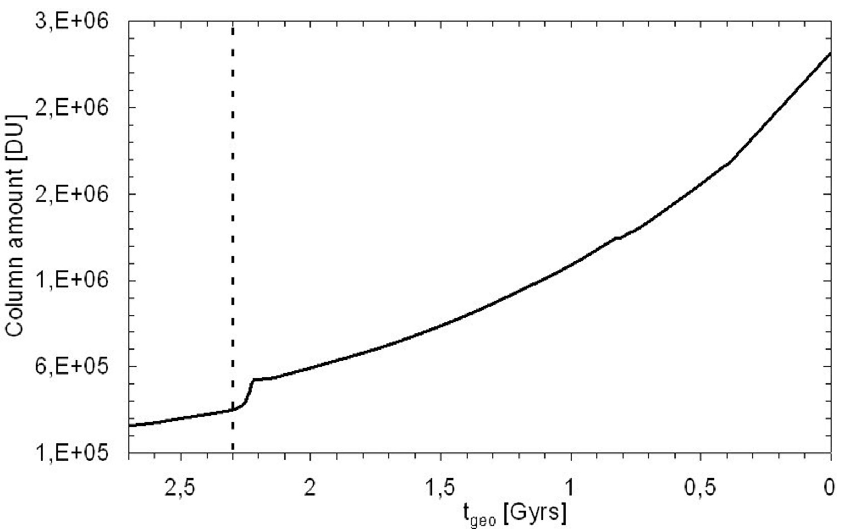} 
  \caption{H$_2$O column amount [DU] as a function of geological time $t_{\rm{geo}}$ for Earth-in-time atmosphere runs considered. The beginning of the GOE at $t_{\rm{geo}}=2.3$ Gyrs is indicated as a vertical dashed line.}
  \label{H2Ocolumnfig} 
\end{figure}

The overall H$_2$O column amount increases on going forwards in geological time as shown in Fig. \ref{H2Ocolumnfig} which can be attributed directly to the temperature behavior shown in Fig. \ref{figure_Tdecrease}. There is a caveat, however, that there could be deviations in our temperature behavior in comparison to the geological record.


\paragraph{\textbf{Nitrous oxide - N$_2$O}}
N$_2$O is an important biosignature. For modern Earth, it is almost exclusively produced by bacteria as part of the (de)nitrifying cycle \citep{ipcc2001}. It is destroyed mainly in the stratosphere via photolysis or via the reaction with excited oxygen (O($^1$D)). 

\begin{figure}
\centering
\includegraphics[width=8.5cm]{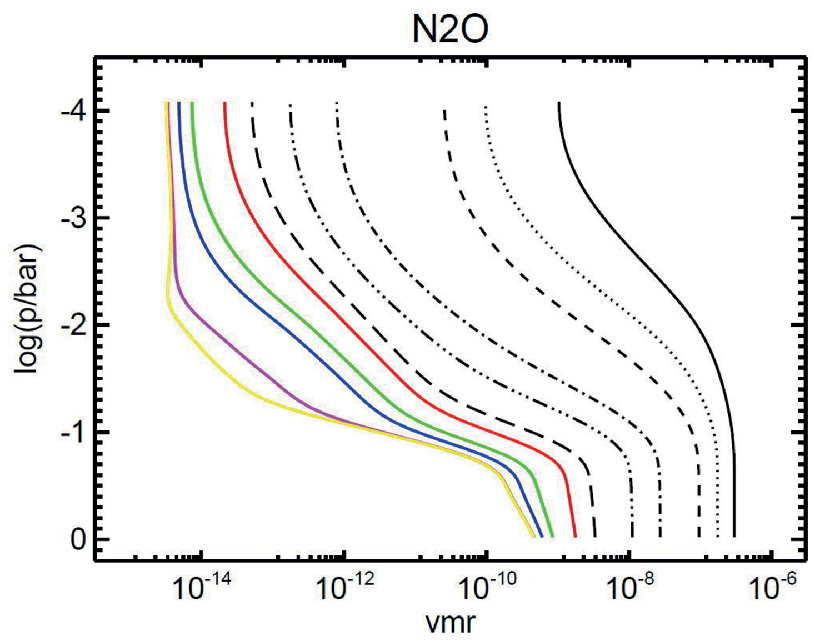} 
  \caption{N$_2$O profiles of Earth-in-time atmosphere runs calculated by the CAB model. Notation as in Fig. \ref{figure_pTvarO2}.}
  \label{N2OvarO2fig}
\end{figure}

\begin{figure}  
\centering
\includegraphics[width=8.5cm]{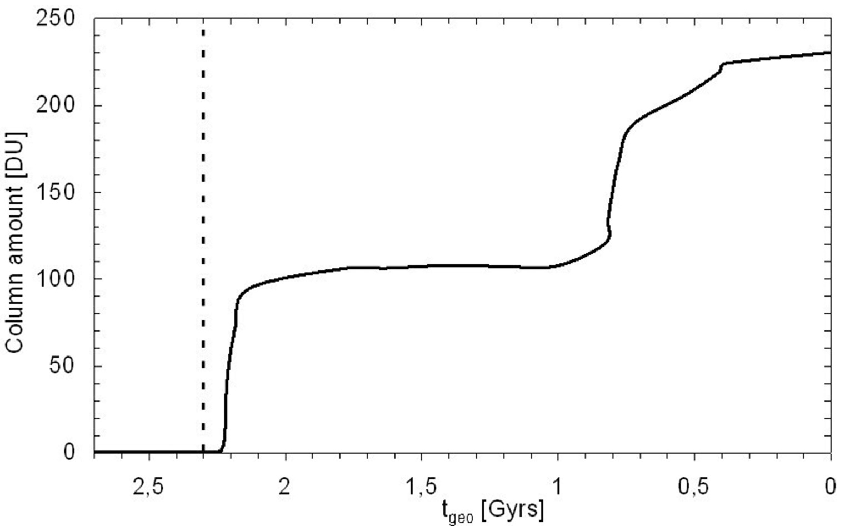} 
  \caption{N$_2$O column amount [DU] as a function of geological time $t_{\rm{geo}}$ for Earth-in-time atmosphere runs considered. The beginning of the GOE at $t_{\rm{geo}}=2.3$ Gyrs is indicated as a vertical dashed line.}
  \label{N2Ocolumnfig}
\end{figure}

\begin{figure}
\centering
\includegraphics[width=8.5cm]{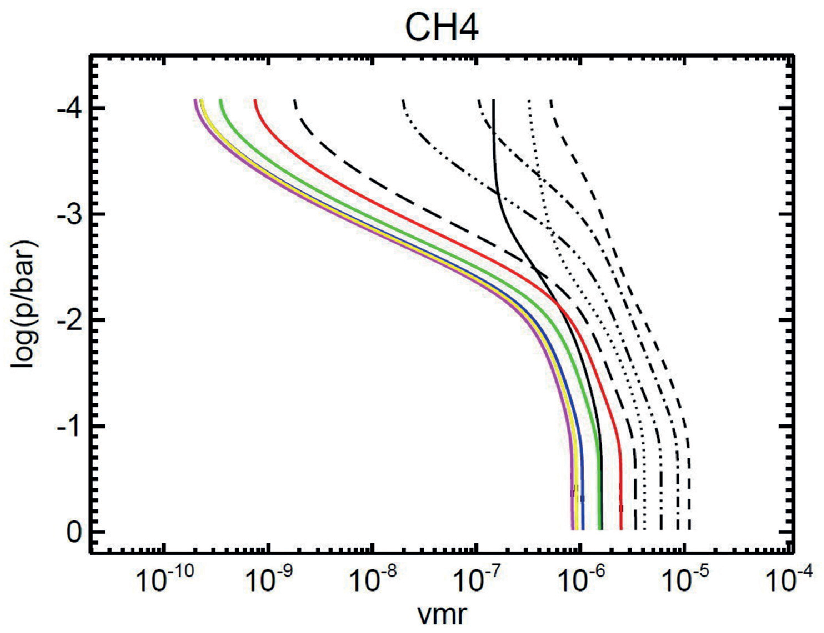} 
  \caption{CH$_4$ profiles of Earth-in-time atmosphere runs calculated by the CAB model. Notation as in Fig. \ref{figure_pTvarO2}.}
  \label{CH4varO2fig}
\end{figure}

Fig. \ref{N2OvarO2fig} shows the N$_2$O vmr profiles for decreased ground level O$_2$ concentrations.
N$_2$O destruction via photolysis becomes more efficient in the atmosphere due to increased UV radiation for atmospheres with lower O$_2$ surface vmrs. This results in decreased N$_2$O concentrations and, hence, lower column amounts (shown in Fig. \ref{N2Ocolumnfig}) in the geological past. In situ inorganic production of N$_2$O is negligibly small.


\paragraph{\textbf{Methane - CH$_4$}}
Atmospheric CH$_4$ is a bioindicator since in addition to biogenic sources some geological sources exist. CH$_4$ is destroyed in the atmosphere mainly by the reaction with hydroxyl (OH) and in the upper stratosphere by photolysis. 

\begin{figure}
\centering
\includegraphics[width=8.5cm]{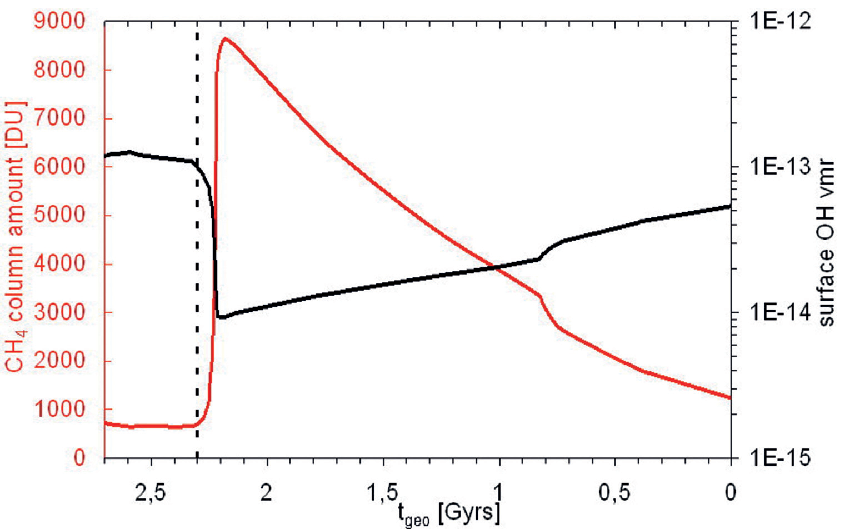} 
  \caption{CH$_4$ column amount [DU] (red) and surface OH vmr (black) as a function of geological time $t_{\rm{geo}}$ for Earth-in-time atmosphere runs considered. The beginning of the GOE at $t_{\rm{geo}}=2.3$ Gyrs is indicated as a vertical dashed line.}
  \label{CH4columnfig}
\end{figure}

Fig. \ref{CH4varO2fig} shows the CH$_4$ profiles for reduced ground level O$_2$ concentration. Note, however, that we fixed the biological surface flux for all scenarios considered. But see also sensitivity run H in Section \ref{sensstudiessection}, which considers a doubled flux. For increasing O$_2$ concentrations, CH$_4$ increases as well, reaches a maximum at 0.1 PAL O$_2$, and then decreases toward higher O$_2$ values. The corresponding column amounts are given in Fig. \ref{CH4columnfig}. The maximum in CH$_4$ column amount at $t_{\rm{geo}}=2.18$ Gyrs (0.1 PAL O$_2$ vmr) is directly linked with a minimum concentration of OH and, hence, minimum destruction of CH$_4$, at the same geological time. For modern Earth, standard OH sources are, for example, H$_2$O + O($^1$D) $\rightarrow$ OH + OH (where O($^1$D) arises mainly from O$_3$ photolysis) and sinks, for example, CO + OH $\rightarrow$ CO$_2$ + H and OH + HO$_2$ $\rightarrow$ H$_2$O + O$_2$. Note that the first sink reaction allegedly destroys OH. Most of the H atoms formed react quickely via the reaction H + O$_2$ + M $\rightarrow$ HO$_2$ + M followed by, for example, HO$_2$ + O$_3$ $\rightarrow$ OH + 2O$_2$, which reforms OH, and thus the reaction between CO and OH does not lead to an efficient OH loss. The second loss reaction is a more permanent sink for OH. 
The main source of OH switches to H$_2$O + h$\nu$ $\rightarrow$ H + OH for the low O$_2$ atmospheric run due to higher UV.  

Fig. \ref{CH4columnfig} illustrates an important result of our work, that is, we calculate an increase in CH$_4$ concentration with increasing O$_2$ vmrs at the GOE.
Previous works \citep [e.g.,][]{claire2006,zahnle2006} with much simpler chemical schemes, however, suggest the opposite effect, namely, that increasing O$_2$ (more oxidizing conditions) would lead to stronger CH$_4$ oxidation (into CO$_2$ and H$_2$O) and, hence, a reduction in CH$_4$. This behavior is hypothesized by the authors to lead to a Snowball Earth.
CH$_4$-oxidation is a complex multireaction process that usually begins by attacking OH on CH$_4$.
OH is photolytically produced, which (generally) means high OH abundances in high UV atmospheres, all else being equal.
In our study, an increase in O$_2$ leads to an increase in O$_3$, which blocks UV. This leads to less OH and, hence, more CH$_4$. Our work suggests that more investigations are required in the chemical feedbacks between CH$_4$ and O$_2$ on the early Earth.
After the GOE, O$_2$ further increases and CH$_4$ is more and more destroyed via the reaction with OH, which then increases because of
higher production via, for example, H$_2$O + O($^1$D) $\rightarrow$ 2OH (since H$_2$O increases as discussed). Note that in our study we assume a constant biogenic input of CH$_4$ of 474 Tg/yr at the surface for every run considered. This value could be rather weak for the early Earth \citep [see][]{claire2006}. Also, due to the toxic nature of O$_2$ to CH$_4$ producing microorganisms, the rise in O$_2$ likely had a severe impact on the biological activity and produced atmospheric CH$_4$, therefore, resulting in a dramatic decrease in biotic CH$_4$ production. This mechanism is not included in the CAB model, and is beyond the scope of this paper.

\paragraph{\textbf{Carbon dioxide - CO$_2$}}
CO$_2$ is an important greenhouse gas that is generally well mixed in the modern Earth atmosphere. 

\begin{figure}
\centering
\includegraphics[width=8.5cm]{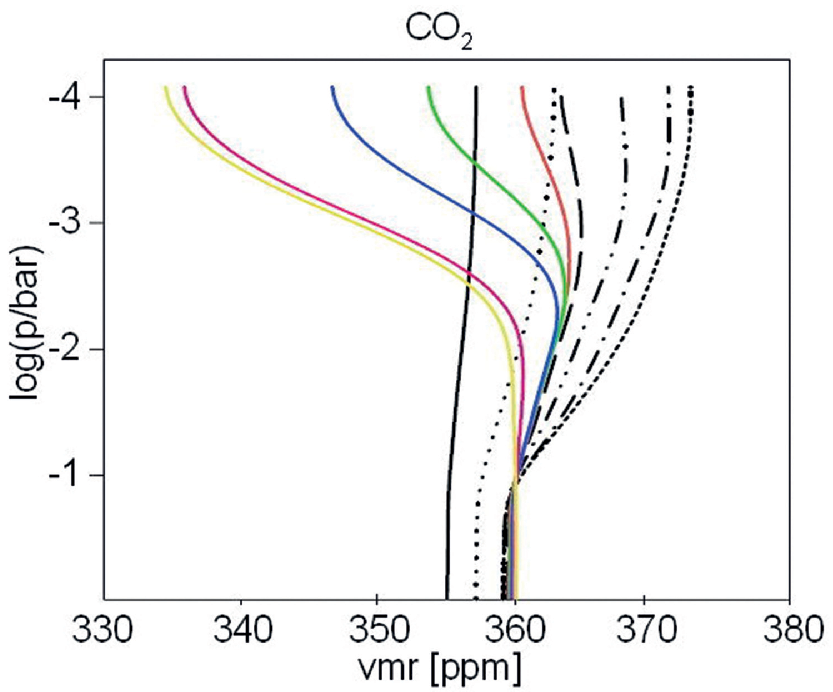} 
  \caption{CO$_2$ profiles of Earth-in-time atmosphere runs calculated by the CAB model. Notation as in Fig. \ref{figure_pTvarO2}.}
  \label{CO2varO2fig}
\end{figure}

\begin{figure}
\centering
\includegraphics[width=8.5cm]{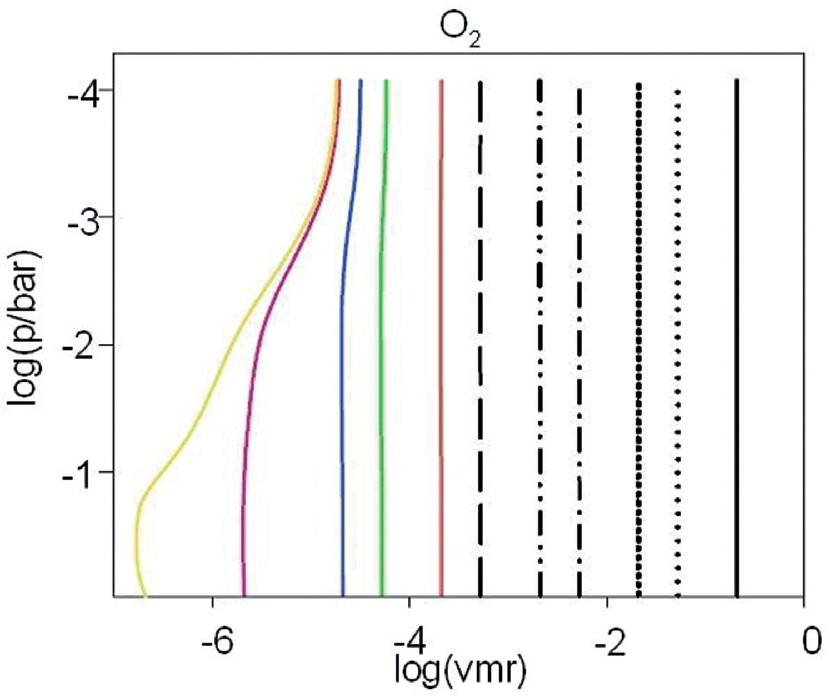} 
  \caption{O$_2$ profiles of Earth-in-time atmosphere runs calculated by the CAB model. Notation as in Fig. \ref{figure_pTvarO2}.}
  \label{O2varO2fig}
\end{figure}

Fig. \ref{CO2varO2fig} shows how the CO$_2$ vmr profiles change on decreasing the ground level O$_2$ concentrations. For decreases down to 0.1 PAL O$_2$, CO$_2$ increases in the stratosphere mainly due to the HO$_{\rm{x}}$ catalyzed net reaction O$_3$ + 3CO $\rightarrow$ 3CO$_2$, which is initialized by the catalyzed photolysis of O$_3$ and O$_2$.
For further decreases in O$_2$, stronger photolysis of CO$_2$ leads to decreased vmrs in the stratosphere. This implies that CO$_2$ does not exhibit an isoprofile throughout the atmosphere for low O$_2$ concentrations. In the case of 10$^{-6}$ PAL O$_2$, the TOA vmr of CO$_2$ is 334 ppm in comparison to the surface vmr of 355 ppm.
There is a caveat, however: CO$_2$ profiles could depend on the rather unconstrained value of the eddy diffusion profile $K(z)$.
 
\paragraph{\textbf{Molecular oxygen - O$_2$}}

Fig. \ref{O2varO2fig} shows that O$_2$ exhibits an isoprofile behavior for atmosphere runs with ground level O$_2$ concentrations above $10^{-4}$ PAL O$_2$, which implies that its concentration is dominated by transport processes rather than chemical production ($P$) and destruction ($L$). For the atmosphere runs with surface O$_2$ concentrations below $10^{-4}$ PAL O$_2$, an anti-correlation between the vertical CO$_2$ and O$_2$ profiles becomes visible implying that the destruction of CO$_2$ provides a source of atmospheric O$_2$ in the upper stratosphere. 
A detailed chemical analysis will be presented in Section \ref{papanalysis} to gain detailed insight into the rather complex production (and destruction) processes for O$_2$. 

\begin{figure}
\centering
\includegraphics[width=8.5cm]{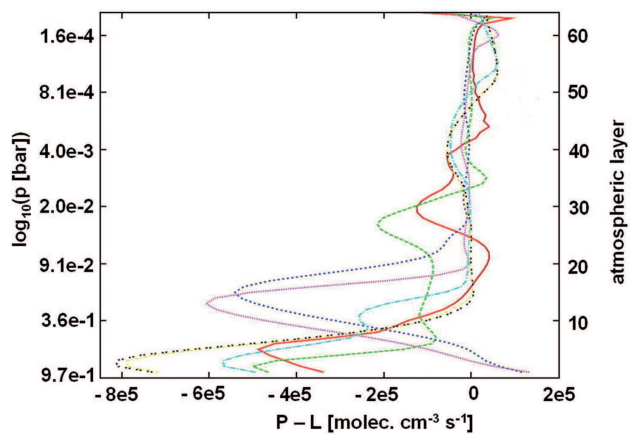} 
  \caption{Atmospheric net $(P-L)$ chemical change of O$_2$ calculated by the CAB model for modern Earth and low O$_2$ atmospheres. Notation: solid red: 1 PAL O$_2$ (modern Earth), dashed green: $10^{-1}$ PAL ($t_{\rm{geo}}=2.18$ Gyrs), dotted dark blue: $10^{-2}$ PAL ($t_{\rm{geo}}=2.22$ Gyrs), dotted purple: $10^{-3}$ PAL ($t_{\rm{geo}}=2.24$ Gyrs), dot-dashed cyan: $10^{-4}$ PAL ($t_{\rm{geo}}=2.28$ Gyrs), dot-dashed yellow: $10^{-5}$ PAL ($t_{\rm{geo}}=2.59$ Gyrs), dot-dot black: $10^{-6}$ PAL ($t_{\rm{geo}}=2.69$ Gyrs).}
  \label{PLO2fig}
\end{figure}

Fig. \ref{PLO2fig} compares the net chemical change 
\begin{equation}
	(\Delta n_i)_{\rm{chem}}=P_i - L_i
\end{equation}
of $i=\rm{O}_2$ in the atmosphere for modern Earth and low O$_2$ atmospheres. Thereby, $n_i$ is the number density of the chemical compound $i$.
For modern Earth O$_2$ is only produced immediately below and above the O$_3$ layer maximum in the stratosphere. 
In general, O$_2$ is mostly destroyed in the atmosphere except for the lowermost atmospheric layers of atmospheres with ground level concentrations between 10$^{-2}$ and  10$^{-3}$ PAL. For very low O$_2$ atmospheres there is a net production at the uppermost layers which is a result of increased CO$_2$ photolysis as mentioned above. 

\begin{figure}
\centering
\includegraphics[width=8.5cm]{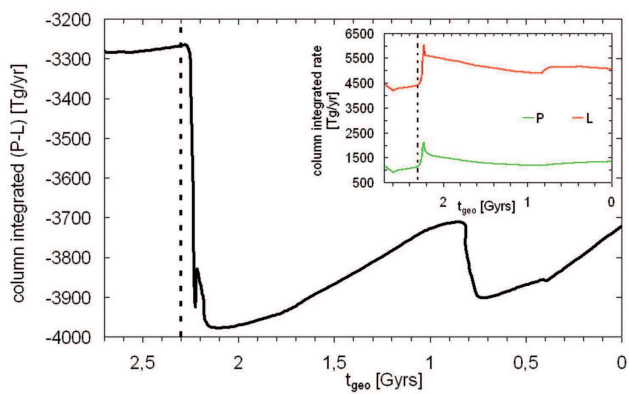} 
  \caption{Column integrated net $(P-L)$ chemical change of atmospheric O$_2$ calculated by the CAB model as a function of geological time $t_{\rm{geo}}$ for Earth-in-time atmosphere runs considered. The beginning of the GOE at $t_{\rm{geo}}=2.3$ Gyrs is indicated as a vertical dashed line. In the upper right panel the column-integrated P (green) and L (red) rates are given as a function of geological time $t_{\rm{geo}}$ for comparison.}
  \label{columnPLfig}
\end{figure}

Fig. \ref{columnPLfig} depicts the net column integrated (global mean) $(P-L)$ chemical change of O$_2$ calculated by the CAB model on going forward in geological time. Addidionally, in the upper right of Fig. \ref{columnPLfig} the column-integrated production and destruction rates of O$_2$ are shown for comparison. This illustrates that O$_2$ is generally destroyed within the atmosphere for all runs considered. Hence, this indicates that a positive O$_2$ flux of the same amount into the atmosphere is required in order to achieve steady-state and maintain O$_2$ levels above zero. It is striking that although the surface O$_2$ concentration varies by several orders of magnitude, the associated O$_2$ surface flux, $\Phi^{\rm{atm}}_{\rm{O}_2}$, into the atmosphere varies between about 3300 and 4000 Tg/yr by only about 20\%. Before the GOE it exhibits an almost constant behavior.
In this atmosphere regime, small changes in the O$_2$ surface flux can support a wide range of different surface O$_2$ vmrs (not shown).
Then, $\Phi^{\rm{atm}}_{\rm{O}_2}$ increases sharply during the GOE as surface O$_2$ concentrations increase strongly. After the GOE, $\Phi^{\rm{atm}}_{\rm{O}_2}$ decreases, followed by a sharp increases at the SOE at 0.8 Gyrs ago and then decreases again. 
Generally, one recognizes that the mathematical solution is not unique, that is, for one atmospheric O$_2$ surface flux there are multiple surface O$_2$ vmrs regarding the different boundary conditions. 
This means that the same magnitude of O$_2$ surface flux (which is a measure of the strength of the chemical production and destruction of O$_2$ in the whole atmosphere) can support either (a) an anoxic atmosphere or (b) a rather oxic atmosphere (which in turn depends on other established factors known to affect O$_2$(g) such as irradiation and volcanic/ metamorphic outgassing). 
The biosphere productivity that corresponds to the O$_2$ surface flux at different ground level O$_2$ concentrations is investigated in Section \ref{biosphere}.


\paragraph{\textbf{Comparison to previous work}}\label{workcomp}
Generally, it is not easy to compare results with those of previous works in the literature regarding the early Earth since either, for example, only stand-alone photochemical models (i.e., without coupled climate calculations) have been presented (e.g. \citealt{kasting1980,kasting1982,kasting1985,zahnle2006}) or coupled climate-chemistry calculations performed using an isoprofile assumption for O$_2$, CO$_2$ and N$_2$ (e.g. \citealt{segura2003}\footnote{\citet{segura2003} did not change the solar constant while reducing surface O$_2$ concentrations, which makes it additionally difficult to compare with.}). Furthermore, previous workers have assumed increased CO$_2$ vmrs (see e.g., \citealt{kasting1984}) and CH$_4$ surface fluxes (see e.g., \citealt{pavlov2003}) to be present in the early Earth atmosphere.

\begin{figure}
\centering
\includegraphics[width=8.5cm]{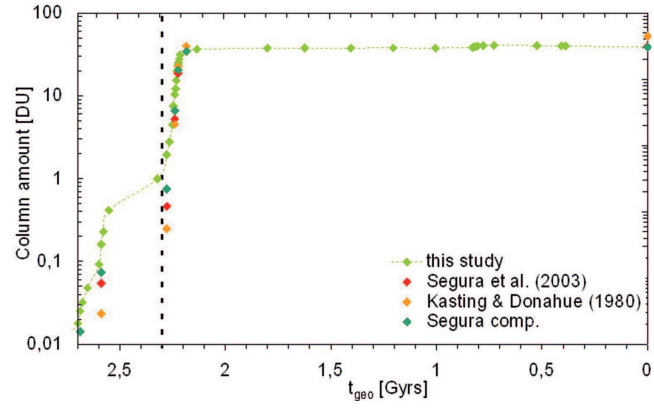} 
  \caption{Column amount of atmospheric O$_3$ in DU as a function of geological time $t_{\rm{geo}}$ for Earth-in-time atmosphere runs considered in comparison to previous work. The beginning of the GOE at $t_{\rm{geo}}=2.3$ Gyrs is indicated as a vertical dashed line.}
  \label{compO3fig}
\end{figure}

Some key chemical reactions resulting in the production and destruction of O$_2$ have, however, also been identified, for example, in \citet{kasting1984} although calculated O$_2$ profiles differ from this work. Focusing on atmospheric O$_3$, Fig. \ref{compO3fig} shows a comparison of the O$_3$ column depth calculated by this study with those of \citet{segura2003} and \citet{kasting1980}. To compare with the work of \citet{segura2003}, additional runs were performed with the CAB model ("Segura comp." shown in dark green) with the initial atmospheric conditions as in the \citet{segura2003} study.
A further important result of this work is shown in Fig. \ref{compO3fig}, where it can be seen that, due to our atmosphere model improvements, we calculate a consistently enhanced O$_3$ column amount compared to previous works in the literature. 
This difference calculated in O$_3$ occurs, for example, due to an increased photolytic destruction of CO$_2$ in the runs of this work leading to production of O$_2$ and, hence, O$_3$. Comparing results of the Segura comp. runs with the results of \citet{segura2003} suggests, for example, a maximum relative change in the O$_3$ column amount of 82\% at $t_{\rm{geo}}=2.28$ Gyrs ($10^{-4}$ PAL O$_2$). For H$_2$O, N$_2$O and CH$_4$ qualitatively similar results were found.


\subsubsection{Pathway Analysis with respect to O$_2$}\label{papanalysis}
For the early Earth analog runs, model calculations generally predict an increase of the O$_2$ vmr with increasing atmospheric height, which implies an in situ atmospheric source of O$_2$ (see Fig. \ref{O2varO2fig}). Furthermore, it can be seen that the CO$_2$ profile is anti-correlated with the O$_2$ profile in the low O$_2$ Earth-in-time atmosphere runs (see Fig. \ref{CO2varO2fig}). This might imply that CO$_2$ serves as a source species for O$_2$. Our goal is to identify how O$_2$ is produced from CO$_2$ in the presence of highly reactive radicals such as OH. We further focus on the O$_2$ destruction pathways to identify the dominant reduced chemical species that are responsible for the consumption of O$_2$.
Therefore, the Pathway Analysis Program (PAP) (for details see Section \ref{papsection}) is applied for an atmosphere with a surface O$_2$ vmr of $10^{-6}$ PAL O$_2$ ($t_{\rm{geo}}=2.688$ Gyrs). Our study represents the first application of PAP in the context of the early Earth. We consider only pathways with an individual contribution larger than 1.5\% of the total production (or destruction) of O$_2$.

In the analysis, we firstly considered all chemical species with an in situ chemical lifetime smaller than that of O$_2$ as a branching point as described in Section \ref{papsection}. It turned out that CO is part of the net reactions of the dominant O$_2$ production and destruction pathways, respectively. This indicates, that for the rates of O$_2$ production and destruction pathways, it is relevant whether CO is treated as a branching point or not (if it is, all CO production and destruction pathways will be combined as far as possible to yield null cycles, i.e., pathways which do not have a net effect on CO; if it is not, CO production and destruction pathways are not combined). For about 40\% of the atmospheric layers considered, CO was used as a branching point.
To ensure a consistent treatment of all O$_2$ production and destruction pathways at every atmospheric height, we repeated the analysis by excluding CO as a branching point.

\paragraph{\textbf{Production pathways and their altitude dependence}}	
Tab. \ref{proPAPpathstab} summarizes the major column-integrated chemical production pathways, P1 - P7, found by PAP and their percentage contributions to the total column-integrated O$_2$ production rate for an atmosphere with a surface O$_2$ vmr of $10^{-6}$ PAL O$_2$. 

The major column-integrated chemical production pathways presented in Tab. \ref{proPAPpathstab} can be categorized into 2 classes as follows:
\begin{itemize}
	\item class PA:\\ O$_2$ is produced by pathways with the net reaction\\ 2CO$_2$ $\rightarrow$ O$_2$ + 2CO
\begin{itemize}
	\item subclass PA1 (P1, P2, P7): catalyzed by HO$_{\rm{x}}$ species,
	\item subclass PA2 (P3, P4): without the presence of HO$_{\rm{x}}$,
\end{itemize}
	\item class PB:\\ O$_2$ is formed by pathways with the net reaction 2O $\rightarrow$ O$_2$
	\begin{itemize}
	\item subclass PB1 (P6): catalyzed by HO$_{\rm{x}}$ species,
	\item subclass PB2 (P5): without the presence of HO$_{\rm{x}}$.
\end{itemize}
\end{itemize}

\begin{figure}
\centering
\includegraphics[width=8.5cm]{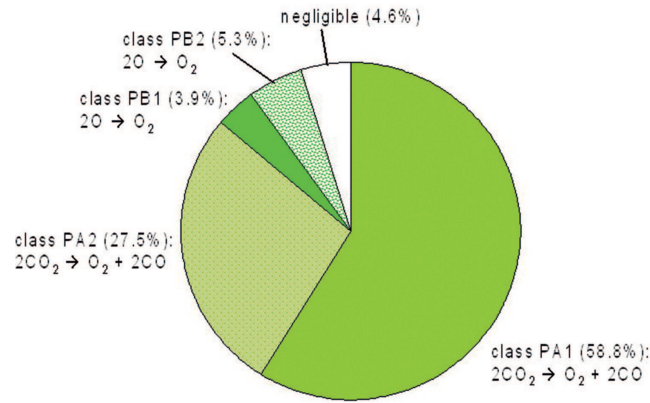}
  \caption{O$_2$ production classes (see Tab. \ref{proPAPpathstab}) and their contributions to the total column-integrated O$_2$ production rate via pathways calculated by PAP for an Earth-like atmosphere with a surface O$_2$ vmr of $10^{-6}$ PAL O$_2$.}
  \label{piePfig}
\end{figure}

The relative contributions of the pathways mentioned above to the total column-integrated production rate are given in Fig. \ref{piePfig}.
Pathways of subclass PA1 and PA2 are initiated by the photolysis of CO$_2$, whereas class PB pathways require overall the presence of atomic oxygen. The main source for O originates from the photolysis of CO$_2$ in the upper stratosphere producing O($^1$D) and, hence, O, which is then transported downwards by eddy diffusion, contributing to enhanced O$_2$ vmrs in the upper stratosphere. Note that at the upper boundary of the photochemistry part of the atmosphere module an effusion flux for O and CO is implemented to simulate this effect from atmospheric layers above the module TOA. Pathway P2 was found by \citet{yung1999} in the context of a pure CO$_2$ dominated atmosphere.

\begin{figure}
\centering
\includegraphics[width=8.5cm]{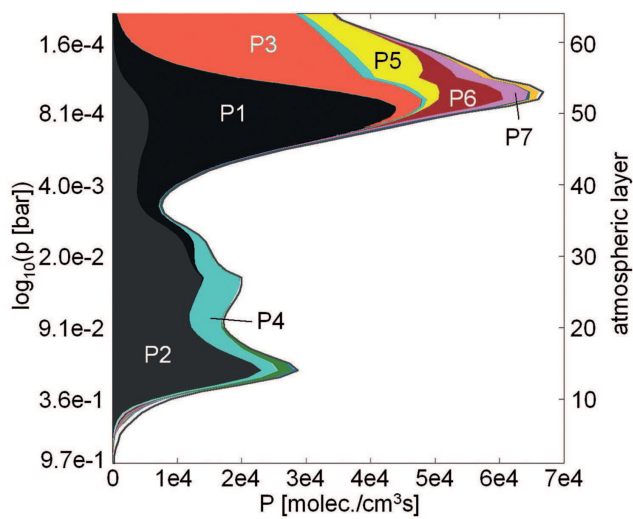} 
  \caption{Altitude dependence of O$_2$ production pathways P1 to P7 (see Tab. \ref{proPAPpathstab}) for an Earth-like atmosphere with a ground level O$_2$ concentration of 10$^{-6}$ PAL.}
  \label{allpro1emin6fig}
\end{figure}

Fig. \ref{allpro1emin6fig} shows the altitude dependence of the production pathways of O$_2$ for an Earth-like atmosphere with a surface O$_2$ vmr of $10^{-6}$ PAL O$_2$, which indicates that 
the production of O$_2$ in the atmosphere is strongest in the upper stratosphere. The total production rate due to chemical pathways calculated by PAP is also shown (solid black line). About 98.7\% of the total column-integrated O$_2$ production rate are explicable via pathways found by PAP with individual contributions above 1.5\%. 
The dominant production pathways P1 and P3 (as well as P7), which produce excited oxygen (O($^1$D)) from CO$_2$ photolysis, act in the upper stratosphere where UV radiation is strongest. 
As the UV radiation passes through the atmosphere, it is absorbed and, hence, the second dominant pathway P2 (and also P4), which involves ground-state oxygen (O) instead of O($^1$D), has its maximum rate at lower altitudes. Due to the enhanced O abundance in the upper stratosphere, which is provided from CO$_2$ photolysis that produces O($^1$D) and is transported downward to the analyzed atmospheric layers, the pathways P5 and P6 have their maximum rate in the upper stratosphere. Note that, except for the photolysis of CO$_2$, pathways P5 and P6 are equal to P4 and P2.

\paragraph{\textbf{Destruction pathways and their altitude dependence}}
Tab. \ref{lossPAPpathstab} summarizes the major column-integrated chemical destruction pathways and their percentage contributions to the total column-integrated O$_2$ destruction rate found by PAP for an atmosphere with a surface O$_2$ vmr of $10^{-6}$ PAL O$_2$. 

These pathways can be categorized into 4 classes:
\begin{itemize}
	\item class LA:\\ O$_2$ is destroyed by pathways with the net reaction O$_2$ + 2CO $\rightarrow$ 2CO$_2$ (CO-oxidation) which are catalyzed by HO$_{\rm{x}}$ and NO$_{\rm{x}}$ species
	\begin{itemize}
	\item subclass LA1 (L1, L2, L6, L8): initialized by the oxidation of H,
	\item subclass LA2 (L3, L5): initiated by the photolysis of O$_2$,
\end{itemize} 
	\item class LB:\\ O$_2$ is consumed by CH$_4$-oxidation pathways
\begin{itemize}
	\item subclass LB1 (L4, L7): net reaction 3O$_2$ + 2CH$_4$ $\rightarrow$ 4H$_2$O + 2CO,
	\item subclass LB2 (L10): net reaction O$_2$ + CH$_4$ $\rightarrow$ H$_2$O + H$_2$ + CO,
\end{itemize} 
	\item class LC (L9):\\ O$_2$ is consumed by pathways with the net reaction O$_2$ + 2H$_2$ $\rightarrow$ 2H$_2$O (H$_2$-oxidation),
	\item class LD (L11):\\ O$_2$ is consumed by pathways with the net reaction O$_2$ + H$_2$O + CO $\rightarrow$ H$_2$O$_2$ + CO$_2$.
\end{itemize}

\begin{figure}
\centering
\includegraphics[width=8.5cm]{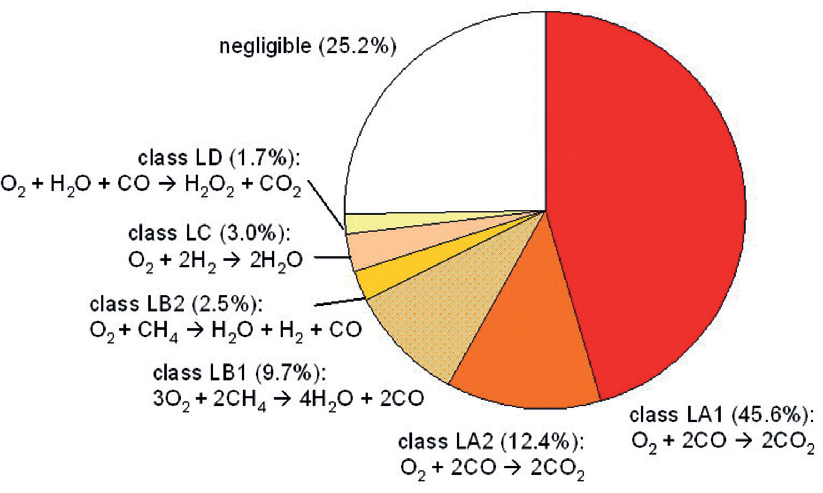} 
  \caption{As for Fig. \ref{piePfig} but for destruction (see also Tab. \ref{lossPAPpathstab}).}
  \label{pieLfig}
\end{figure}

Their relative contributions to the total destruction rate via pathways are given in Fig. \ref{pieLfig}. 
Generally, O$_2$ is consumed forming CO$_2$, CO, H$_2$, H$_2$O, and H$_2$O$_2$.
Pathways of class LA lead to the oxidation of CO mainly via HO$_{\rm{x}}$. Thereby, in the subclass LA2 pathways are initiated by the photolysis of O$_2$. The complex pathways of the subclasses LB1 and LB2 destroy O$_2$ via the oxidation of CH$_4$, whereas class LC leads to the oxidation of H$_2$. Furthermore, there is a small contribution from class LD that leads to the formation of H$_2$O$_2$ via oxidation of CO and H$_2$O. 
This is rather surprising since H$_2$O$_2$ is a highly reactive (oxidizing) tropospheric species (produced by self-reaction of HO$_2$ and removed via photolysis and fast surface deposition). H$_2$O$_2$ features a lifetime of typically a few hours and is therefore set to be a branching point species throughout the troposphere (see Section \ref{papsection}). Nevertheless, L11 suggests a pathway in which H$_2$O$_2$ is overall produced. We interpret this result as follows: PAP was supplied with input from rate data for chemical reactions that occur only (in situ) in the atmosphere. It was not supplied with rates for other processes such as dry and wet deposition. Since H$_2$O$_2$ features a rapid depositional surface sink, and therefore to attain an equilibrium concentration for this species in the atmospheric model, there must exist a pathway in the atmosphere that overall forms a net source of H$_2$O$_2$, to balance the strong surface depositional sink. It is this net source that PAP has found in the form of L11.

Several pathways shown in Tab. \ref{lossPAPpathstab} have also been found by previous workers in the context of the martian atmospheric chemistry, for example, L1 by \citet{parkinson1973} and \citet{yung1999}, L2 by \citet{stock2012a}, L3 by \citet{mcelroy1972}, L6 by \citet{nair1994} and \citet{yung1999}, and L8 by \citet{sonnemann2006}. This implies that pathways that are important for the CO$_2$-dominated martian atmosphere may also play a key role in an Earth-like atmosphere with low O$_2$ and 1 PAL CO$_2$ abundances.
%

\begin{figure}
\centering
\includegraphics[width=8.5cm]{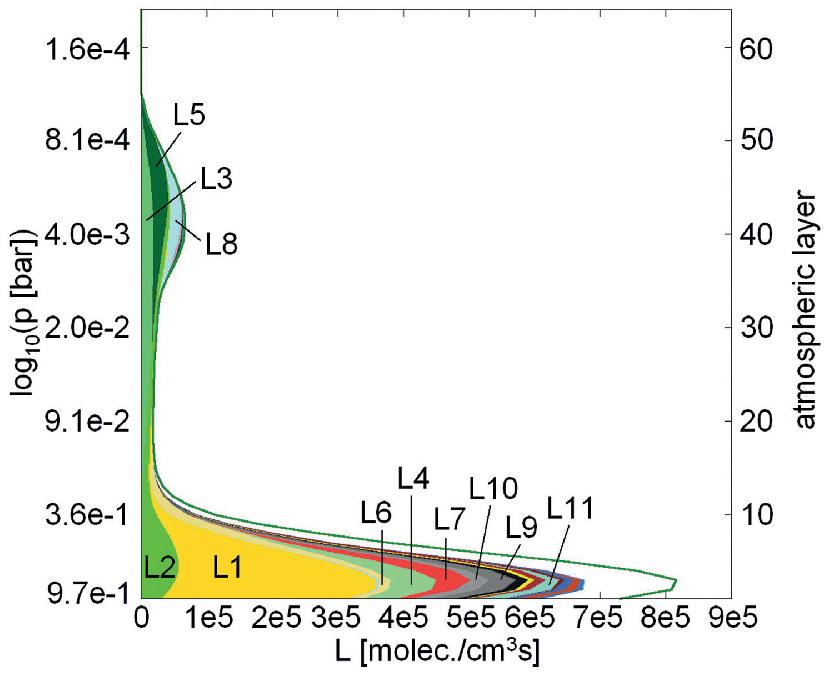} 
  \caption{As for Fig. \ref{allpro1emin6fig} but for destruction (see also Tab. \ref{lossPAPpathstab}).}
  \label{alldes1emin6fig}
\end{figure}

Fig. \ref{alldes1emin6fig} shows the altitude dependence of the O$_2$-consumption pathways for an Earth-like atmosphere with a surface O$_2$ vmr of $10^{-6}$ PAL O$_2$. The total destruction rate due to chemical pathways calculated by PAP is also shown (solid green line). About 84.1\% of the total column-integrated O$_2$ destruction rate is due to pathways with individual contributions above 1.5\%.  Especially, in the troposphere the destruction of O$_2$ is composed of a large number of pathways, which individually contribute less than 1.5\% to the total column-integrated destruction rate.
Fig. \ref{alldes1emin6fig} suggests two regimes for the destruction of O$_2$ in the atmosphere as follows:
\begin{itemize}
	\item  The main consumption of O$_2$ in the troposphere occurs via the oxidation of reduced gases such as CO, CH$_4$, and H$_2$ (but only to a very small amount ($<1.5$\%) by reduced sulfur such as H$_2$S) and is composed of a large number of different pathways. The dominant pathway L1 leads to CO-oxidation catalyzed by HO$_{\rm{x}}$ (CO is provided by diffusion from the upper atmosphere where it is strongly produced from CO$_2$ photolysis), whereas the complex CH$_4$ oxidation pathways L4, L7, and L10 have smaller contributions. Note that our results may change depending of the CH$_4$ concentration and, hence, biological CH$_4$ flux inserted at the surface.
	The minor pathway L9 results in the oxidation of H$_2$, which is delivered by volcanic emissions from the surface and via diffusion from the upper atmosphere. H$_2$ is destroyed in situ in the lower atmosphere. There is an additional small contribution from pathway 11 that results in the oxidation of CO in the presence of H$_2$O.
		\item  A minor contribution (L3, L5, and L8) to the destruction pathways originates in the stratosphere. At these altitudes O$_2$ consumption during CO oxidation is most efficient due to a maximum in HO$_2$ abundance (not shown), whereby pathways L3 and L5 are initiated by the photolysis of O$_2$.
\end{itemize}

\hfill\\
In summary, for an Earth-like atmosphere with low surface O$_2$ concentrations ($10^{-6}$ PAL O$_2$) atmospheric O$_2$ is produced in the upper stratosphere mainly from the in situ photolytic destruction of CO$_2$ (whereas a small contribution relies on the delivery of O by atmospheric diffusion).
O$_2$ is mainly destroyed in the lower atmosphere by complex pathways resulting mainly in the formation of H$_2$O and CO$_2$ via the oxidation of CO, CH$_4$, and H$_2$. Thereby, there is a minor contribution from pathways resulting in the formation of H$_2$ and H$_2$O$_2$. In both the production and destruction of O$_2$, CO plays a key role. Note that the abiotic in situ production rate of O$_2$ in the atmosphere from CO$_2$ photolysis as well as the O$_2$ destruction rate are comparatively smaller than the O$_2$ flux originating from the surface biosphere.


\subsection{Biogeochemical Modeling}\label{biosphere}
The CAB model calculates how much input from oxygenic photosynthesis, $N$ (NPP), is needed to maintain the surface flux $\Phi^{atm}_{\rm{O}_2}$ calculated by the atmospheric chemistry module for a given surface vmr of O$_2$.

\begin{figure}
	\centering
\includegraphics[width=8.5cm]{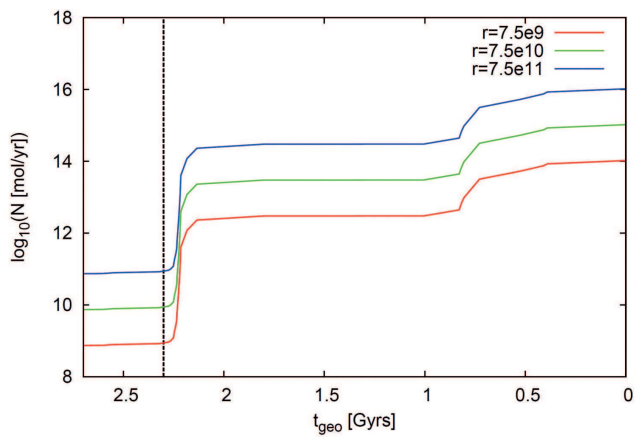} 
  \caption{Calculated input from oxygenic photosynthesis, $N$, in mol O$_2$ per year for a constant value of anoxygenic photosynthesis, $r$, as a function of geological time $t_{\rm{geo}}$: $r=7.5\cdot10^{10}$ mol O$_2$ equiv./yr is the present value \citep{holland2006} (green), $r=7.5\cdot10^{11}$ mol O$_2$ equiv./yr is an upper limit for the Archean \citep{holland2006} (blue) and for comparison calculations were performed for a constant value of $r=7.5\cdot10^{9}$ mol O$_2$ equiv./yr (red). The beginning of the GOE at $t_{\rm{geo}}=2.3$ Gyrs is indicated as a vertical dashed line.}
  \label{N_rconstfig}
	\end{figure}

Assuming three constant inputs from anoxygenic photosynthesis ($r=7.5\cdot10^{9}$, $7.5\cdot10^{10}$ (present value, \citealt{holland2006}), and $7.5\cdot10^{11}$ mol O$_2$ equiv./yr (Archean upper limit, \citealt{holland2006})), Fig. \ref{N_rconstfig} shows the resulting $N$ as a function of geological time $t_{\rm{geo}}$. 

Before the GOE at about 2.3 Gyrs $N$ exhibits an almost constant behavior, although O$_2$ concentrations vary over two orders of magnitude (see Tab. \ref{runtable}). After the GOE, it increases modestly with increasing O$_2$ concentrations and does not reflect the sometimes oscillating behavior found for the atmospheric O$_2$ surface flux (see $(P-L)$ chemical change behavior in Section \ref{photochem} for comparison). 
On inserting Eq. \ref{fluxstag} into Eq. \ref{2fluxes} one obtains
\begin{equation}\label{eq1}
	\Phi_{\rm{O}_2}^{\rm{atm}}=\rm{v}_p(\rm{O}_2)([\rm{O}_2]^{\rm{aq}}-H_{\rm{O}_2} p_{\rm{O}_2})C\;\;.
\end{equation}
Eq. \ref{eq1} can now be rearranged for $[\rm{O}_2]^{\rm{aq}}$, which is calculated by the biogeochemical module (see Eq. \ref{steadymasterbgc} and \ref{aquO2}) and directly depends on $N$. For $N$, two regimes can then be identified as follows:
\begin{itemize}
	\item Regime I (low O$_2$ regime before the GOE):  $H_{\rm{O}_2} p_{\rm{O}_2}\ll \frac{\Phi_{\rm{O}_2}^{\rm{atm}}}{C \rm{v}_p(\rm{O}_2)}$:\\
	Eq. \ref{eq1} simplifies to $[\rm{O}_2]^{\rm{aq}}\simeq\frac{\Phi_{\rm{O}_2}^{\rm{atm}}}{C\rm{v}_p(\rm{O}_2)}$. Since in this regime $\Phi_{\rm{O}_2}^{\rm{atm}}\approx$ const. $[\rm{O}_2]^{\rm{aq}}$ is also approximately constant and, hence, $N$ (which determines $[\rm{O}_2]^{\rm{aq}}$) exhibits a constant behavior as well.
	
	\item Regime II (high O$_2$ regime after the GOE):  $H_{\rm{O}_2} p_{\rm{O}_2}\gg \frac{\Phi_{\rm{O}_2}^{\rm{atm}}}{C \rm{v}_p(\rm{O}_2)}$:\\
	The effect of changes in the O$_2$ surface flux, $\Phi_{\rm{O}_2}^{\rm{atm}}$, calculated by the atmospheric chemistry module upon $N$, is small compared to the effect of the O$_2$ partial pressure, $p_{\rm{O}_2}$. On increasing the ground level O$_2$ concentration the O$_2$ partial pressure (but also the total surface pressure which is determined by the evaporation of H$_2$O and, hence, surface temperatures, not shown) increases resulting in higher values for $[\rm{O}_2]^{\rm{aq}}$; hence $N$ increases as well in order to satisfy Eq. \ref{eq1}.
	This implies that the input from oxygenic photosynthesis is not sensitive to
	 atmospheric chemistry which governs $\Phi_{\rm{O}_2}^{\rm{atm}}$ needed to maintain a certain atmospheric O$_2$ surface vmr. 
\end{itemize}

\noindent The switch between both regimes can be found where $H_{\rm{O}_2} p_{\rm{O}_2}=\frac{\Phi_{\rm{O}_2}^{\rm{atm}}}{C\rm{v}_p(\rm{O}_2)}$ and occurs at about $4\cdot10^{-4}$ PAL O$_2$ which corresponds to about $t_{\rm{geo}}=2.25$ Gyrs ago..

\begin{figure}
	\centering
\includegraphics[width=8.5cm]{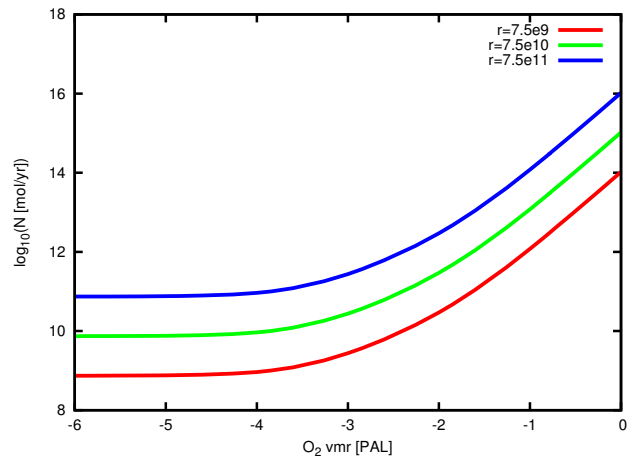} 
  \caption{\textcolor{red}{As for Fig. \ref{N_rconstfig} but over O$_2$ vmr instead of geological time $t_{\rm{geo}}$.}}
  \label{N_rconstfigoverO2}
	\end{figure}

Results in Fig. \ref{N_rconstfig} can also be plotted over O$_2$ mixing ratio, see Fig. \ref{N_rconstfigoverO2} suggesting that one level of net primary productivity from oxygenic photosynthesis, $N$, supports only one specific O$_2$ surface vmr. Please note, that the y-axis in Fig. \ref{N_rconstfigoverO2} is logarithmic and in the case of low O$_2$ atmospheres, changes in $N$ are very small but monotonically increasing.

\subsubsection{Modern Earth}			
Regarding the NPP from oxygenic photosynthesis, $N$, the CAB model calculates $N_{\rm{CAB}}=1.1$ Pmol O$_2$/yr for modern Earth at 1 PAL O$_2$, which is within a factor of 4 of the rather uncertain observed present value ($N_{\rm{obs}}=3.75$ Pmol O$_2$/yr) (see red curve in Fig. \ref{N_rconstfig}). The calculated organic carbon burial rate of $2.1$ Tmol O$_2$/yr is in the range of the rather uncertain observed value of 10 Tmol O$_2$/yr \citep{holland1978}, whereas the calculated organic carbon reservoir of $3.5\cdot10^{20}$ mol is somewhat less than the present observed value of $1.66\cdot10^{21}$ mol \citep{holland1978}. Since our study assumes steady-state, the organic carbon burial rate, $\beta(N + r)$, is equal to the weathering rate of organic carbon $wC^{\rm{org}}$.

\subsubsection{Archean Earth}
Modern-day oceanic NPP mostly originates at the coastlines on continental shelves because the abundances of nutrients are strongly increased due to river input. However, due to the lack of knowledge about the length of the continental boundary over time, we used the continental area instead to estimate the Archean NPP, although this assumption might be crude. If we assume that the continental area was reduced (see \citet{taylor1991} for evolution models of the continental crust over Earth's history) and the burial efficiency was probably higher in the past, then the NPP would have been reduced to 1\% of the modern-day value. This estimate is calculated by assuming the following: 
\begin{itemize}
	\item only 3\% of the surface of the Earth was covered by continents at 2.45 Gyrs ago (before the GOE) \citep{pesonen2003,taylor1991} which equals about 10\% of today's continental coverage. If one assumes that the NPP is proportional to the continental area then this leads to a reduction of NPP by a factor of 10.
	\item the absolute organic carbon burial rate of about 10 Tmol/yr is rather constant over time because the fraction of volcanic carbon that is buried as organic carbon (about 20\%) has not changed strongly since at least 3.0 Gyrs (see \citet{schidlowski1983} but also \citet{Shields2002} for carbon isotope data from the Precambrian). Modern burial effciency of organic carbon in the ocean is about 0.2\% \citep{berner1982,betts1991} resulting in modern NPP of 5 Pmol/yr. Since the Archean burial efficiency was likely similar to that of the modern, euxinic Black Sea of about 2\% \citep{arthur1994}, the NPP should have been further decreased by a factor of 10.
\end{itemize}

For the time of the Fe deposition found from the Hamersley BIF (2.69 - 2.44 Gyr ago), the blue curve in Fig. \ref{N_rconstfig} has to be considered, yielding a value of $N_{\rm{CAB}}=8.4\cdot10^{10}$ mol O$_2$/yr which is less than the estimated value of $N_{\rm{archean}}=0.01\cdot N_{\rm{obs}}= 3.75\cdot10^{13}$ mol O$_2$/yr that was derived for the discussion points above. This large discrepancy is likely related to uncertainties, for example, in ocean circulation, length of continental shelves, etc. Unfortunately, along with these uncertainties, burial efficiency is highly indeterminate for the Archean. Furthermore, note that there exist differences between our model output and the proxy data. In the case of the O$_2$ concentrations, we use the upper limits (for sulfur MIF, detrital siderite) or the lower limits (for beggiota, animals, charcoal). Some initial test runs (not shown) varying the O$_2$ concentration in the Archean by a factor of 10 to 100 affected the NPP by only about 10\%.

\subsubsection{Comparison to Previous Work}
It is difficult to compare the results of this study to previous box-model work by, for example, \citet{claire2006} and \citet{goldblatt2006}, because they do not consider a full coupled atmospheric column model with climate and photochemistry. In \citet{claire2006} and in the transient runs by \citet{goldblatt2006}, the rise of atmospheric O$_2$ is accompanied by a collapse in atmospheric CH$_4$ due to the O$_2$-CH$_4$ feedback as already described in Section \ref{photochem}.

Furthermore, \citet{goldblatt2006} found a bistability in atmospheric O$_2$ as a result of the non-linear behavior of their CH$_4$ oxidation paramerization. In their Fig. 2b, a certain range in input from oxygenic photosynthesis for a constant $r$ can support two different atmospheric O$_2$ concentrations, for example, a low oxygen state (before the GOE) and a high oxygen state (after the GOE) due to the O$_2$-O$_3$-UV feedback.
However, such bistability behavior is not observed in our work since we treat the oxidation of CH$_4$ as part of the atmosphere modeling, which is decoupled from the biogeochemical modeling. It was shown (see Figs. \ref{N_rconstfig} and \ref{N_rconstfigoverO2}) that, for atmospheres before $t_{\rm{geo}}=2.25$ Gyrs ago (with surface O$_2$ concentrations below $10^{-4}$ PAL), rather small changes in $N$ result in different O$_2$ vmrs, and hence the atmosphere-surface system is very sensitive to changes in input from oxygenic photosynthesis.


\section{Sensitivity Results}\label{sensstudiessection}
\begin{figure}
\centering
  \includegraphics[width=8.5cm]{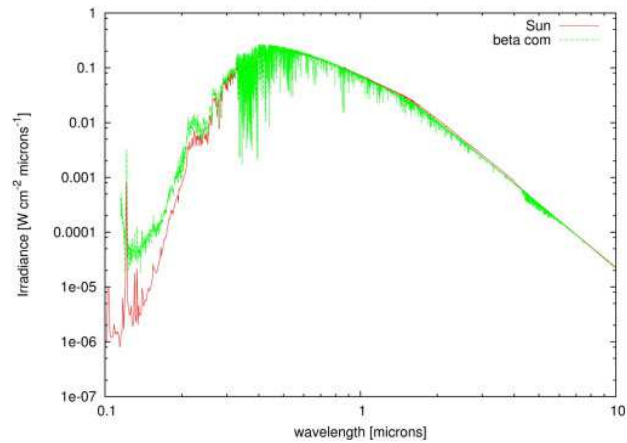} 
  \caption{\textcolor{red}{Normalized stellar spectrum of $\beta$ Com (green) in comparison to the spectrum of the Sun (red) by \citet{Gueymard2004}.}}\label{sunbetacom}
\end{figure}

Sensitivity studies were performed to derive the key parameters and boundary conditions that have the strongest impact upon the behavior of the O$_2$ surface flux calculated by the atmospheric module of the CAB model and the corresponding biosphere response.
The following scenarios were considered:
\begin{itemize}
\item \textit{Scenario A}: here the assumed evolutionary path of O$_2$ is set to be constant to 10$^{-5}$ O$_2$ vmr before 2.5 Gyrs ago. $S(t)$ and volcanic and metamorphic outgassing remain as for the control scenario;
	\item crustal mineral redox buffer: 
		\begin{itemize} 
			\item \textit{Scenario B}: more reduced: $\Delta f_{\rm{O}_2}=-1$ (QFM-1),
			\item \textit{Scenario C}: more oxidized: $\Delta f_{\rm{O}_2}=+1$ (QFM+1); 
  	\end{itemize}
  \item heat flow:
  	\begin{itemize}
			\item \textit{Scenario D}: $2Q$, 
			\item \textit{Scenario E}: $0.5Q$;
		\end{itemize}					
  \item \textit{Scenario F}: eddy diffusion profile is set constant to $K(z)=10^5$ cm$^2$ $s^{-1}$ \citep{kuhn1979} \citep [control run: $K(z)$ from][]{massie1981};
  \item \textit{Scenario G}: surface CO$_2$ concentration: 5 PAL CO$_2$ (sediment data place an upper limit of 10-100 PAL, see \citealt{rye1995,sheldon2006,hessler2004} but see also \citealt{rosing2010} for a more conservative limit);
  \item \textit{Scenario H}: doubled CH$_4$ surface flux (control scenario: 474 Tg CH$_4$/yr)
  \item \textit{Scenario I}: incident stellar radiation: young sun analog star $\beta$ Com (see Fig. \ref{sunbetacom} for comparison of stellar spectra of $\beta$ Com and the Sun)
	\item \textit{Scenario J}: Pasteur point: lowered to $10^{-5}$ PAL O$_2$ \citep{stolper2010};
	\item \textit{Scenario K}: solubility of O$_2$ in ocean water:\\
	Henry's law constant $H_{\rm{O}_2} = 9.8\cdot 10^{-4}$ M/atm (Broecker and Peng, 1982) at 25$^{\circ}$C, from the diffusion coefficient $K_{\rm{diffusion,O_2}}=2.26\cdot10^{-5}$ cm$^2$/s at 25$^{\circ}$C (Broecker and Peng, 1974) one can calculate a piston velocity as $\rm{v}_p(\rm{O}_2)=5.65\cdot 10^{-3}$ cm s$^{-1}$.	
\end{itemize}

In the case of Scenario A (not shown), only a minor increase in O$_2$ surface flux before 2.5 Gyrs ago compared to the control scenario (with a maximum relative change of 0.6\%) is found. Hence atmospheric chemistry is not strongly affected compared to the scenario discussed in the previous sections (control run). The effect upon the biosphere response is a little larger. To maintain a constant O$_2$ vmr of $10^{-5}$ at the surface, $N$ increases by about 12 \% compared to the control scenario because of the slightly enhanced O$_2$ surface flux.

\begin{figure}
	\centering
\includegraphics[width=8.5cm]{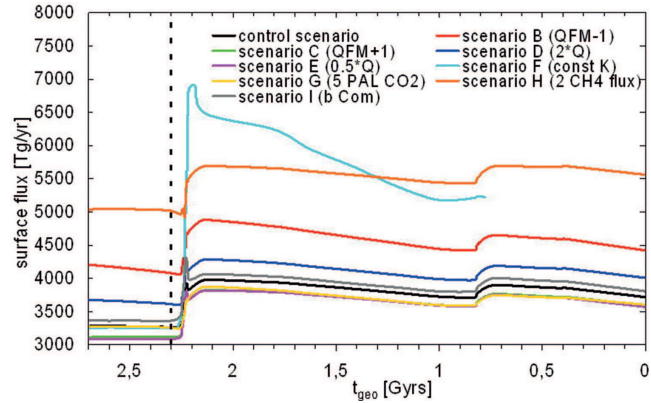}
  \caption{Calculated atmospheric O$_2$ surface flux as a function of geological time $t_{\rm{geo}}$ for a number of different parameter studies compared to the control scenario described in Section \ref{scenariodescription}: scenario B (QFM-1), scenario C (QFM+1), scenario D ($2Q$), scenario E ($0.5Q$), scenario F (const. $K(z)$), scenario G (5 PAL CO$_2$), scenario H (2 CH$_4$ flux) and scenario I ($\beta$ Com). The parameter studies concerning the Pasteur point (scenario J) and the ocean solubility (scenario K) have the same atmospheric O$_2$ surface flux as the control scenario. The beginning of the GOE at $t_{\rm{geo}}=2.3$ Gyrs is indicated as a vertical dashed line.}
  \label{O2flux_sensfig}
\end{figure}

Fig. \ref{O2flux_sensfig} shows how the atmospheric O$_2$ surface flux responds for the other sensitivity runs.	A larger O$_2$ surface flux is needed to maintain the O$_2$ surface vmr in the cases of scenario B (QFM-1), scenario D ($2Q$), and scenario H (doubled CH$_4$ flux) because of the increased amount of reduced gases emitted at the surface. 

In the case of scenario F (const. eddy diffusion), the constant $K(z)$-profile leads to a stronger O$_3$ layer (not shown) with an O$_3$ column amount of about 430 DU at $t_{\rm{geo}}=2.18$ Gyrs ago (0.1 PAL O$_2$ - cf. 263 DU for control run). Furthermore, the O$_3$ surface vmr of $1.2\cdot10^{-7}$ is higher than that in the control scenario ($2.61\cdot10^{-8}$) (not shown). 
At 0.1 PAL O$_2$, eddy diffusion is greatly increased compared to modern Earth above 10 km, which leads to a more efficient transport of chemical species toward lower pressure, and hence species such as CH$_4$ and N$_2$O are more quickly removed. HO$_{\rm{x}}$ and NO$_{\rm{x}}$ are weaker, resulting in enhanced O$_3$ abundances.
Hence, the overall increased O$_3$ abundances result in increased O$_2$ surface fluxes after the GOE because the production of O$_3$ is an effective sink for O$_2$. In the case of scenario I ($\beta$ Com), a slightly increased O$_2$ surface flux is calculated, which is also a direct consequence of the increased O$_3$ abundance in such an atmosphere. Due to enhanced UV radiation emitted by the star below about 300 nm, the production of O$_3$ via the Chapman mechanism, which represents a loss process for O$_2$, is stimulated. The O$_2$ surface flux is generally lower for scenario C (QFM+1) and scenario E ($0.5Q$) because less reduced chemical species are emitted into the atmosphere. In the case of scenario G (5 PAL CO$_2$,) the O$_2$ fluxes at the surface are less than that in the control scenario after the GOE, which implies that increased CO$_2$ concentrations lead to higher production of O$_2$ in the atmosphere because CO$_2$ photolysis is a source for O and O($^1$D) and, hence, O$_2$. Before the GOE, O$_2$ fluxes are equal to the control scenario. 
 
\begin{figure}
	\centering
\includegraphics[width=8.5cm]{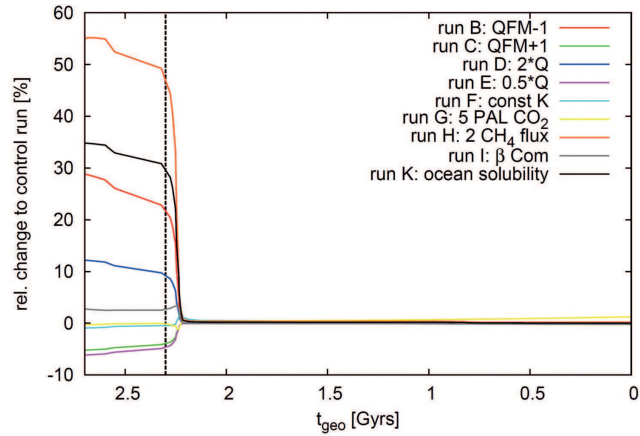}
  \caption{Calculated relative change in \% of the input from oxygenic photosynthesis, $N$, for the different parameter changes compared to the control scenario as a function of geological time $t_{\rm{geo}}$. Scenario J is not shown in this figure. The beginning of the GOE at $t_{\rm{geo}}=2.3$ Gyrs is indicated as a vertical dashed line.}
  \label{N_sens_allfig}
\end{figure}

The effect upon the biosphere is shown in Fig. \ref{N_sens_allfig}. 
For the different scenarios we find: 
\begin{itemize}
	\item Scenario B (QFM-1, red): Due to the overall, more reduced atmospheric conditions and the associated destruction of O$_2$, a higher O$_2$ surface flux is required, and hence a higher $N$ is needed, which results in a rel. change of about 29\% at 2.7 Gyrs ago; 
	\item Scenario C (QFM+1, green dashed): More oxidized conditions due to less reduced gases emitted at the surface lead to a maximum decrease of 5 \% in $N$;
	\item Scenario D ($2Q$, blue dotted): A doubled heat flow results in more reduced gases and, hence, an overall increase in $N$ by 12 \%; 
	\item Scenario E ($0.5Q$, magenta): A smaller heat flow yields a decrease in $N$ by about 6\%;
	\item Scenario F (const. $K(z)$, cyan): There are only minor relative changes with a maximum of 2-3\% at 2.22 Gyrs ago; 
	\item Scenario G (5 PAL CO$_2$, yellow): Minor relative changes are observed with a minimum value of 2\% at 2.24 Gyrs; 
	after the GOE, the relative change increases to 1-2\%, which reflects the higher surface pressures and temperatures compared to the control run; 
	\item Scenario H (2 CH$_4$ flux, orange): A doubled CH$_4$ flux at the surface has a strong effect on the biosphere response and results in a maximum relative change of 55 \% in $N$; 
	\item Scenario I ($\beta$ Com, grey): A higher UV radiation below 300 nm results in a larger O$_3$ column and, hence, a higher O$_2$ surface flux and therefore an increase in $N$ by about 4\%; 
	\item Scenario K: (ocean solubility, green): To maintain the same O$_2$ flux as for the control run, the aqueous concentration $[\rm{O}_2]^{\rm{aq}}$ and, hence, $N$ must increase because the solubility constant and piston velocity of O$_2$ are lower (see Eq. \ref{fluxstag}), a maximum increase of 35\% is observed;
	\item Scenario J (Pasteur point, not shown in Fig. \ref{N_sens_allfig}):	In the case of the reduction of the Pasteur point, O$_2$ can be consumed by aerobic respiration down to lower aqueous O$_2$ concentrations which results in a stronger O$_2$ loss, and therefore the input from oxygenic photosynthesis must increase by several orders of magnitude (10$^3$). Hence, better knowledge of the Pasteur point is important.
	 
\end{itemize}

Generally speaking, relative changes in $N$ (except for the Pasteur point run) are very small after the GOE, although the corresponding O$_2$ surface fluxes calculated by the atmospheric chemistry module are increased or decreased in comparison to the control scenario (see Fig. \ref{O2flux_sensfig}). This is because, in the regime after the GOE, it is the partial pressure of O$_2$, and hence surface conditions ($p_{\rm{surf}}$, $T_{\rm{surf}}$), rather than the O$_2$ surface flux reflecting the chemical behavior of O$_2$ in the atmosphere (as discussed in Section \ref{biosphere}) which is responsible for the biosphere response.


\section{Conclusion}\label{resultsection}
This is the first study to our knowledge that includes the coupling between geological, biological, and (detailed) atmospheric climate and photochemistry processes. By applying our unique global-mean atmospheric column
module with interactive O$_2$, CO$_2$, and N$_2$ (CAB model) which can respond to
detailed atmospheric chemistry and radiative transfer, our study investigated consistently chemical and biological sources and sinks that controll the evolution of atmospheric O$_2$ on Earth and potentially on Earth-like extrasolar planets.
Our new model reasonably reproduces observations of the modern Earth for both atmospheric data (atmospheric chemical species, climate) and biological-geological data (net primary productivity and burial).
Our results suggest that, in the atmosphere, CO$_2$ (we assume modern-day surface values) is not an isoprofile (as is commonly assumed in the literature) in the middle atmosphere for all early Earth analog runs, and O$_2$ is no longer an isoprofile for runs with less than $10^{-4}$ O$_2$ surface vmrs. 

The O$_2$ bistability found by \citet{goldblatt2006} is not observed in our calculations due to our consistent atmospheric treatment of CH$_4$ oxidation. Furthermore, we calculate an increase in CH$_4$ concentration with increasing O$_2$ vmrs during the GOE in contrast to previous works \citep [e.g.,][]{claire2006,zahnle2006} with much simpler chemical schemes, which have suggested the opposite effect. Important caveats, however, include uncertainties, for example, in the modeled temperature evolution due to lacking knowledge about the evolution of greenhouse gas evolution (CO$_2$, CH$_4$, N$_2$O) as well as missing processes in the model, for example, the response of methanogen activity to changing atmospheric O$_2$(g) and temperature. 

For the first time, we applied the Pathway Analysis Program for O$_2$ in the early Earth analog atmosphere and found that photolysis of CO$_2$ is an important abiotic source of O$_2$ (and CO) in the upper atmosphere. However, this process is not able to change the overall atmospheric O$_2$ content on early Earth analog planets. In the early reducing atmosphere, O$_2$ is mainly destroyed via oxidation of CO, CH$_4$, and H$_2$ in the lower atmosphere. It was found that CO is a crucial chemical species in chemical pathways that produce and destroy O$_2$.

On applying the CAB model to the early Earth analog atmospheres, we calculated O$_2$ surface fluxes over the geological timescales. An important result was that different types of atmospheres, that is, with different oxidizing natures, could be supported by one O$_2$ surface flux respectively. More work is required to understand the nature of the oxidizing-reducing feedbacks in the early Earth analog atmospheres. But due to the coupling of biogeochemical processes to the atmosphere we found that there is only one net primary productivity from oxygenic photosynthesis that is needed to maintain the O$_2$ surface vmr for every type of atmosphere considered. The atmospheric O$_2$ surface flux is most sensitive to changes in the biogenic CH$_4$ surface emission and a crustal chemical redox buffer that emits more reduced than oxidized gases \mbox{(QFM-1)}, but also to the assumed eddy diffusion profile. Atmospheres with enhanced CO$_2$ concentrations lead to a decrease in the required atmospheric O$_2$ surface flux. The net primary productivity of the biosphere is strongly influenced by changes in the Pasteur point, biogenic CH$_4$ surface emission, ocean solubility, and crustal chemical redox buffers that emit more reduced gases (QFM-1), and hence resolving the uncertainties in these quantities are crucial for a better understanding of early Earth analog atmospheres. 

Although our main focus was on early Earth analog planets it is nevertheless instructive to briefly discuss our results in an Earth-like exoplanetary context. Clearly this discussion remains hypothetical and requires future work.
Regarding Earth-like extrasolar planets, results suggest that different atmospheric "states" of O$_2$ could exist even for similar biomass output. The PAP analysis suggests that atmospheres where volatile organic compounds build up atmospheric O$_2$ are disfavored due to oxidation reactions. Furthermore, extrasolar Earth-like planets with strong interior outgassing of reducing gases can lead to so-called false negatives for O$_2$ (e.g., PAP results show that CO and/or CH$_4$ containing pathways destroy O$_2$ and, hence, mask the existence of O$_2$ producing biospheres). Since all O$_2$ destruction pathways found feature HO$_{\rm{x}}$, steamy high UV exoatmospheres could also lead to similar false negatives in O$_2$. Extrasolar Earth-like planets with strong mixing (e.g. due to a weak temperature inversion) have a potentially strong positive effect on atmospheric O$_3$. Varying key parameters (e.g., interior outgassing) in low O$_2$ containing exoatmospheres ($<1\%$ PAL) significantly impacts the net productivity of a potential biosphere for a given surface O$_2$ concentration.

\newpage

\noindent \textbf{Acknowledgments:}\\
This research was supported by the \emph{Helmholtz Gemeinschaft}
  through the research alliance "Planetary Evolution and Life".
  
  This study has received financial support from the French State in the frame of the "Investments for the future" Programme IdEx Bordeaux, reference ANR-10-IDEX-03-02.
  
  M. Godolt acknowledges funding by the Helmholtz Foundation via the Helmholtz Postdoc Project "Atmospheric dynamics and photochemistry of Super-Earth planets".
  
  \added[]{This work was partially funded by grant AyA 2012–32237 awarded by the
Spanish Ministerio 353 de Economia y Competitividad.}

  \added[]{The authors thank the unknown referee for his/her valuable comments.}

\noindent \textbf{Author Disclosure Statement:}\\
No competing financial interests exist.

\bibliographystyle{agufull08}
\bibliography{Bibo_1D}

\newpage 

\begin{longtable} {c c c}

\caption{Present-day fluxes of volcanic gases considered in the photochemistry module. Fluxes are given in Teragrams (Tg ($10^{12}$ g))/yr.}\\
\hline
 species i & flux $P_{\rm{volc,i}}^{\rm{now}}$ [Tg/yr] & reference\\
\hline 
\endfirsthead
\hline	
 species i & flux $P_{\rm{volc,i}}^{\rm{now}}$ [Tg/yr] & reference\\
\hline 
\endhead
\hline
\endfoot
\hline
\endlastfoot

		 H$_2$  & 9.6 &  \citet{holland2002}\\ 
		 H$_2$S & 1.4 &  \citet{halmer2002}\\
		 SO$_2$ & 15  &  \citet{halmer2002}\\
		 CO     & 1.5 &  \citet{zahnle2006}\\
		 CH$_4$ & 4   &  \citet{kvenvolden2005}\\ 
		 CO$_2$ & 367 &  \citet{morner2002} 
\label{tab1}
\end{longtable}

\newpage 

\begin{longtable} {c c}
\caption{Processes of the marine biosphere after \citet{goldblatt2006}. Organic carbon is represented by CH$_2$O.}\\
\hline
 Process name & Reaction\\
\hline 
\endfirsthead
\hline	
 Process name & Reaction\\
\hline 
\endhead
\hline
\endfoot
\hline
\endlastfoot

		  anoxygenic & 4Fe$^{2+}$ + CO$_2$ + 11H$_2$ + h$\nu$ $\rightarrow$ \\ 
		   photosynthesis & 4Fe(OH)$_3$ + CH$_2$O + 8H$^+$ \\\hline
		  oxygenic photosynthesis & CO$_2$ +H$_2$O + h$\nu$ $\rightarrow$ CH$_2$O + O$_2$\\\hline
		  aerobic respiration &  CH$_2$O + O$_2$ $\rightarrow$ CO$_2$ + H$_2$O\\\hline
		  acetotrophic methanogenic&  2CH$_2$O $\rightarrow$ CH$_3$COOH\\ 
		  fermentation &  CH$_3$COOH $\rightarrow$ CH$_4$ + CO$_2$\\\hline
		  methanotrophic & CH$_4$ + 2 O$_2$ $\rightarrow$ CO$_2$ + 2H$_2$O \\ 
		  anaerobic respiration& 
\label{biospheretab}
\end{longtable}

\newpage 

\begin{longtable} {c c}
\caption{Relative change of relevant chemical species profiles against RG2011 over altitude.}\\
\hline
species & relative change $X$ [\%]\\
\hline 
\endfirsthead
\hline	
species & relative change $X$ [\%]\\
\hline 
\endhead
\hline
\endfoot
\hline
\endlastfoot

		 O$_3$  & -0.7 to 3.5 \\ 
		 H$_2$O & -2.2 to 0.3 \\
		 CH$_4$ & 0.3 to 1.2  \\
		 N$_2$O     & -1.7 to 0.2 \\
		 CH$_3$Cl &  0.0 to 2.7 \\ 
		 CO$_2$ & 0.0 to 0.6 
\label{table:2}
\end{longtable}

\newpage 

\begin{longtable} {c c c c}
\caption{Column amount [DU] (Dobson unit, $2.69\cdot 10^{16}$ molecules cm$^{-2}$) of selected chemical species for modern Earth calculated by the CAB model and RG2011.}\\
\hline
species  & CAB model         & RG2011            & relative change $X$ [\%]\\
\hline 
\endfirsthead
\hline	
species  & CAB model         & RG2011            & relative change $X$ [\%]\\
\hline 
\endhead
\hline
\endfoot
\hline
\endlastfoot

		 O$_3$    & 305.0             &  304.7             & 0.1  \\ 
		 H$_2$O   & $2.418\cdot 10^6$ &  $2.428\cdot 10^6$ & -0.4 \\
		 CH$_4$   & 1237              &  1233              & 0.3  \\
		 N$_2$O   & 230.5             &  231.3             & -0.3 \\
		 CH$_3$Cl & 0.358             &  0.358             &  0.0 
\label{table:3} 
\end{longtable}

\newpage 

\begin{longtable} {c c c}

\caption{Assumed O$_2$ evolution in this work for early Earth analog atmospheres. The surface O$_2$ concentration is derived from the possible evolutionary path of O$_2$ by \citet{catling2005} (see Fig. \ref{figure_sauerstoff}). The geological time $t_{\rm{geo}}$ is interpolated from O$_2(t_{\rm{geo}})$.}\\
\hline
 eon & $t_{\rm{geo}}$ [Gyrs] & surface O$_2$ [PAL]\\
\hline 
\endfirsthead
\hline	
 eon & $t_{\rm{geo}}$ [Gyrs] & surface O$_2$ [PAL]\\
\hline 
\endhead
\hline
\endfoot
\hline
\endlastfoot

  Phanerozoic&0.0   	& 1.0\\
  					&0.39  & $9.05\cdot10^{-1}$   \\
  					&0.41  & $8.50\cdot10^{-1}$   \\
  					&0.52  & $7.14\cdot10^{-1}$   \\\hline 
  Paleozoic & 0.73 & $5.50\cdot10^{-1}$   \\
  					&0.78  & $4.10\cdot10^{-1}$   \\
  					&0.81  & $3.00\cdot10^{-1}$   \\
  					&0.82  & $2.50\cdot10^{-1}$   \\
  					&0.83  & $2.00\cdot10^{-1}$   \\
  					&1.01  & $1.64\cdot10^{-1}$   \\
  					&1.21  & $1.64\cdot10^{-1}$   \\
  					&1.41  & $1.64\cdot10^{-1}$   \\
 					  &1.63  & $1.64\cdot10^{-1}$   \\
  					&1.80  & $1.64\cdot10^{-1}$   \\
  					&2.13  & $1.43\cdot10^{-1}$   \\
  					&2.18  & $1.00\cdot10^{-1}$   \\
  					&2.21  & $5.50\cdot10^{-2}$   \\
  					&2.22  & $2.50\cdot10^{-2}$   \\
  					&2.22  & $1.43\cdot10^{-2}$   \\
  					&2.22  & $1.00\cdot10^{-2}$   \\
  					&2.23  & $5.50\cdot10^{-3}$   \\
  					&2.23  & $2.50\cdot10^{-3}$   \\
  					&2.24  & $1.43\cdot10^{-3}$   \\
  					&2.24  & $1.00\cdot10^{-3}$   \\
  					&2.25  & $5.50\cdot10^{-4}$   \\
  					&2.25  & $2.50\cdot10^{-4}$   \\
  					&2.27  & $1.43\cdot10^{-4}$   \\
  					&2.28  & $1.00\cdot10^{-4}$   \\
  					&2.32  & $5.50\cdot10^{-5}$   \\\hline
  	Archean &2.55  & $2.50\cdot10^{-5}$   \\
  					&2.58  & $1.43\cdot10^{-5}$   \\
  					&2.59  & $1.00\cdot10^{-5}$   \\
  					&2.60  & $5.50\cdot10^{-6}$   \\
  					&2.65   &	$2.50\cdot10^{-6}$   \\
  					&2.68   & $1.43\cdot10^{-6}$   \\
  					&2.69   & $1.00\cdot10^{-6}$   \\
  					&2.70   & $5.50\cdot10^{-7}$   \\
  					&2.75   & $2.50\cdot10^{-7}$   
\label{runtable}
\end{longtable}

\newpage 

\begin{longtable} {c c c c c}

\caption{Summary of dominant chemical production (P) pathways of O$_2$ found by PAP and their percentage contribution to the total column-integrated O$_2$ production rate for an Earth-like atmosphere with a surface O$_2$ vmr of $10^{-6}$ PAL O$_2$. Only pathways that contribute $>1.5$\% of the total column-integrated production rate are shown. For explanation of classes and subclasses see text.}\\
\hline
class & subclass & pathway number & contribution $[\%]$ & pathway\\
\hline 
\endfirsthead
\hline	
class & subclass & pathway number & contribution $[\%]$ & pathway\\
\hline 
\endhead
\hline
\endfoot
\hline
\endlastfoot


 PA & PA1 & P1 & 28.4 & 2$\cdot$(CO$_2$ + h$\nu$ $\rightarrow$ CO + O($^1$D))\\			
  & &	& & 2$\cdot$(O($^1$D) + N$_2$ $\rightarrow$ O + N$_2$)\\			
  & &	& & HO$_2$ + O $\rightarrow$ OH + O$_2$\\			
  & &	& & OH + O $\rightarrow$ H + O$_2$\\			
  & &	& & H + O$_2$ + M $\rightarrow$ HO$_2$ + M\\			
  & &	& & \textbf{net: 2CO$_2$ $\rightarrow$ O$_2$ + 2CO}\\\hline
  	
 & PA1 & P2 & 28.1 & 2$\cdot$(CO$_2$ + h$\nu$ $\rightarrow$ CO + O)\\			
 &  &	& & HO$_2$ + O $\rightarrow$ OH + O$_2$\\			
 &  &	& & OH + O $\rightarrow$ H + O$_2$\\			
  & &	& & H + O$_2$ + M $\rightarrow$ HO$_2$ + M\\			
  & &	& & \textbf{net: 2CO$_2$ $\rightarrow$ O$_2$ + 2CO}\\\hline
  	
 & PA1 & P7 & 2.3 & 2$\cdot$(CO$_2$ + h$\nu$ $\rightarrow$ CO + O($^1$D))\\			
	&  & & & 2$\cdot$(O($^1$D) + N$_2$ $\rightarrow$ O + N$_2$)\\
 &  &	& & O + O$_2$ + M $\rightarrow$ O$_3$ + M\\			
  & &	& & OH + O $\rightarrow$ H + O$_2$\\			
  & &	& & H + O$_3$ $\rightarrow$ OH + O$_2$\\			
  & &	& & \textbf{net: 2CO$_2$ $\rightarrow$ O$_2$ + 2CO}\\\hline
  	
 & PA2 & P3 & 20.7 & 2$\cdot$(CO$_2$ + h$\nu$ $\rightarrow$ CO + O($^1$D))\\			
  & &	& & 2$\cdot$(O($^1$D) + N$_2$ $\rightarrow$ O + N$_2$)\\			
  & &	& & O + O + M $\rightarrow$ O$_2$ + M\\			
  & &	& & \textbf{net: 2CO$_2$ $\rightarrow$ O$_2$ + 2CO}\\\hline
  	
 & PA2 & P4 & 6.8 & 2$\cdot$(CO$_2$ + h$\nu$ $\rightarrow$ CO + O)\\
  & &  & & O + O + M $\rightarrow$ O$_2$ + M\\			
  & &	& & \textbf{net: 2CO$_2$ $\rightarrow$ O$_2$ + 2CO}\\\hline 
  	 	  	
 PB & PB1 & P6 & 3.9 & OH + O $\rightarrow$ H + O$_2$\\					
  & &	& & H + O$_2$ + M $\rightarrow$ HO$_2$ + M\\					
  & &	& & HO$_2$ + O $\rightarrow$ OH + O$_2$\\		
  & &	& & \textbf{net: 2O $\rightarrow$ O$_2$}\\\hline
  
 & PB2 & P5 & 5.3 & O + O + M $\rightarrow$ O$_2$ + M\\			
  & &	& & \textbf{net: 2O $\rightarrow$ O$_2$}
\label{proPAPpathstab}
\end{longtable}

\newpage 

\begin{longtable} {c c c c c}

\caption{As for Tab. \ref{proPAPpathstab} but for O$_2$ destruction (L) pathways.}\\
\hline
class & subclass & pathway number & contribution $[\%]$ & pathway\\
\hline 
\endfirsthead
\hline	
class & subclass & pathway number & contribution $[\%]$ & pathway\\
\hline 
\endhead
\hline
\endfoot
\hline
\endlastfoot


 LA & LA1 & L1	 & 28.5 & 2$\cdot$(H + O$_2$ + M $\rightarrow$ HO$_2$ + M)\\			
    & & & & HO$_2$ + HO$_2$ $\rightarrow$ H$_2$O$_2$ + O$_2$\\			
    & & & & H$_2$O$_2$ + h$\nu$ $\rightarrow$ OH + OH\\			
    & & & & 2$\cdot$(CO + OH $\rightarrow$ CO$_2$ + H)\\			
  	& & & & \textbf{net: O$_2$ + 2CO $\rightarrow$ 2CO$_2$}\\\hline

 & LA1 & L2 & 9.4 & 2$\cdot$(H + O$_2$ + M $\rightarrow$ HO$_2$ + M)\\			
  &  & & & HO$_2$ + O $\rightarrow$ OH + O$_2$\\			
   & & & & HO$_2$ + h$\nu$ $\rightarrow$ OH + O\\			
   & & & & 2$\cdot$(CO + OH $\rightarrow$ CO$_2$ + H)\\			
  & &	& & \textbf{net: O$_2$ + 2CO $\rightarrow$ 2CO$_2$}\\\hline
  
  & LA1 & L6 & 4.4 & 2$\cdot$(H + O$_2$ + M $\rightarrow$ HO$_2$ + M)\\
   & & & & HO$_2$ + O $\rightarrow$ OH + O$_2$\\
   & & & & NO + HO$_2$ $\rightarrow$	NO$_2$ + OH\\	
   & & & & 2$\cdot$(CO + OH $\rightarrow$ CO$_2$ + H)\\	
   & & & & NO$_2$ + h$\nu$ $\rightarrow$ NO + O\\					
   & & & & \textbf{net: O$_2$ + 2CO $\rightarrow$ 2CO$_2$}\\\hline
  	
 & LA1 &  L8 & 3.4 & H + O$_2$ + M $\rightarrow$ HO$_2$ + M\\
   & & & & H + HO$_2$ $\rightarrow$ OH + OH\\			
   & & & & 2$\cdot$(CO + OH $\rightarrow$ CO$_2$ + H)\\					
  & &	& & \textbf{net: O$_2$ + 2CO $\rightarrow$ 2CO$_2$}\\\hline  
    	
 & LA2 & L3 & 7.4 & O$_2$ + h$\nu$ $\rightarrow$ O + O\\			
   & & & & 2$\cdot$(HO$_2$ + O $\rightarrow$ OH + O$_2$)\\			
   & & & & 2$\cdot$(CO + OH $\rightarrow$ CO$_2$ + H)\\			
   & & & & 2$\cdot$(H + O$_2$ + M $\rightarrow$ HO$_2$ + M)\\			
  & &	& & \textbf{net: O$_2$ + 2CO $\rightarrow$ 2CO$_2$}\\\hline

 & LA2 &  L5 & 5.0 & O$_2$ + h$\nu$ $\rightarrow$ O + O($^1$D)\\
  & &  & & O($^1$D) + N$_2$ $\rightarrow$ O + N$_2$\\			
   & & & & 2$\cdot$(HO$_2$ + O $\rightarrow$ OH + O$_2$)\\			
   & & & & 2$\cdot$(CO + OH $\rightarrow$ CO$_2$ + H)\\			
   & & & & 2$\cdot$(H + O$_2$ + M $\rightarrow$ HO$_2$ + M)\\			
  & &	& & \textbf{net: O$_2$ + 2CO $\rightarrow$ 2CO$_2$}\\\hline
  
   LB & LB1 & L4 & 5.4 & H$_2$O$_2$ + h$\nu$ $\rightarrow$ OH + OH\\			
  & & & & 2$\cdot$(CH$_4$ + OH $\rightarrow$ CH$_3$ + H$_2$O)\\	
  & & & & 2$\cdot$(CH$_3$ + O$_2$ + M $\rightarrow$ CH$_3$O2 + M)\\	
  & & & & 2$\cdot$(CH$_3$O$_2$ + HO$_2$ $\rightarrow$ CH$_3$OOH + O$_2$)\\	
  & & & & 2$\cdot$(CH$_3$OOH + h$\nu$ $\rightarrow$ H$_3$CO + OH)\\	
  & & & & 2$\cdot$(H$_2$CO + OH $\rightarrow$ H$_2$O + HCO)\\	
  & & & & 2$\cdot$(HCO + O$_2$ $\rightarrow$ HO$_2$ + CO)\\	
  & & & & HO$_2$ + HO$_2$ $\rightarrow$ H$_2$O$_2$ + O$_2$\\	
  & & & & 2$\cdot$(H$_3$CO + O$_2$ $\rightarrow$ H$_2$CO + HO$_2$)\\	
  & &	& & \textbf{net: 3O$_2$ + 2CH$_4$ $\rightarrow$ 4H$_2$O + 2CO}\\\hline
  & LB1 & L7 & 4.3 & 2$\cdot$(H$_2$CO + OH $\rightarrow$ H$_2$O + HCO)\\			
  & & & & 2$\cdot$(HCO + O$_2$ $\rightarrow$ HO$_2$ + CO)\\	
  & & & & 3$\cdot$(HO$_2$ + HO$_2$ $\rightarrow$ H$_2$O$_2$ + O$_2$)\\
  & & & & 3$\cdot$(H$_2$O$_2$ + h$\nu$ $\rightarrow$ OH + OH)\\		
  & & & & 2$\cdot$(CH$_4$ + OH  $\rightarrow$  CH$_3$ + H$_2$O)\\
  & & & & 2$\cdot$(CH$_3$ + O$_2$ + M $\rightarrow$ CH$_3$O$_2$ + M)\\		
  & & & & 2$\cdot$(CH$_3$O$_2$ + OH $\rightarrow$ H$_3$CO + HO$_2$)\\
  & & & & 2$\cdot$(H$_3$CO + O$_2$ $\rightarrow$ H$_2$CO + HO$_2$)\\	
  & &	& & \textbf{net: 3O$_2$ + 2CH$_4$ $\rightarrow$ 4H$_2$O + 2CO}\\\hline 
   
& LB2 & L10 & 2.5 & CH$_3$OOH + h$\nu$ $\rightarrow$ H$_3$CO + OH\\				
  &  & & & CH$_4$ + OH  $\rightarrow$  CH$_3$ + H$_2$O\\
  &  & & & CH$_3$ + O$_2$ + M $\rightarrow$ CH$_3$O$_2$ + M\\
   & & & & CH$_3$O$_2$ + HO$_2$ $\rightarrow$ CH$_3$OOH + O$_2$\\	
   & & & & H$_3$CO + O$_2$ $\rightarrow$ H$_2$CO + HO$_2$\\	
   & & & & H$_2$CO + h$\nu$ $\rightarrow$ H$_2$ + CO\\
   & & & & \textbf{net: O$_2$ + CH$_4$ $\rightarrow$ H$_2$O + H$_2$ + CO}\\\hline
   
LC & & L9 & 3.0 & 2$\cdot$(H + O$_2$ + M $\rightarrow$ HO$_2$ + M)\\
   & & & & HO$_2$ + HO$_2$ $\rightarrow$ H$_2$O$_2$ + O$_2$\\
   & & & & H$_2$O$_2$ + h$\nu$ $\rightarrow$	OH + OH\\		
   & & & & 2$\cdot$(H$_2$ + OH $\rightarrow$ H$_2$O + H)\\					
  & &	& & \textbf{net: O$_2$ + 2H$_2$ $\rightarrow$ 2H$_2$O}\\\hline 
    
LD & & L11 & 1.7 & CO + OH $\rightarrow$ CO$_2$ + H\\
   & & & & H$_2$O + h$\nu$ $\rightarrow$ H + OH\\
   & & & & 2$\cdot$(H + O$_2$ + M $\rightarrow$ HO$_2$ + M)\\		
   & & & & HO$_2$ + HO$_2$ $\rightarrow$ H$_2$O$_2$ + O$_2$\\					
  & &	& & \textbf{net: O$_2$ + H$_2$O + CO $\rightarrow$ H$_2$O$_2$ + CO$_2$}
\label{lossPAPpathstab}
\end{longtable}
    
\newpage    
    
\end{document}